\documentclass{wiley-article}

\usepackage{siunitx} 
\usepackage{url,hyperref}
\usepackage{booktabs}
\usepackage{subcaption}
\usepackage{color, colortbl}
\definecolor{Gray}{gray}{0.9}
\usepackage{CJKutf8}
\colorlet{shades}{red!15!green!10!blue!15}
\colorlet{links}{red!40!green!200!blue!20}

\newcommand{\shade}{\colorbox{shades}} 

\newcommand{\chainlet}[2]{\mathbb{C}_{#1 \rightarrow #2}}  
\newcommand{\vol}[1]{\mathbb{A}(#1)}  
\newcommand{\occ}[1]{\mathbb{O}(#1)} 
\usepackage{csquotes}
\usepackage{subcaption}
\newcommand{\green}[1]{{#1}}
\newtheorem{info}{Info}

\papertype{Original Article}
\paperfield{Journal Section}

\title{Blockchain Networks: Data Structures of Bitcoin, Monero, Zcash, Ethereum, Ripple and Iota}

\author[1]{Cuneyt Gurcan Akcora}
\author[2]{Yulia R. Gel}
\author[3]{Murat Kantarcioglu}

\affil[1]{Departments of Statistics and Computer Science, University of Manitoba, Winnipeg, MB R3T-2N2, Canada}
\affil[2]{Department of Mathematical Sciences, University of Texas at Dallas, Richardson, TX75080, USA}
\affil[3]{Department of Computer Science, University of Texas at Dallas, Richardson, TX75080, USA}

\corraddress{Cuneyt Gurcan Akcora, Departments of Computer Science and Statistics, University of Manitoba, Winnipeg, MB R3T-2N2, Canada}
\corremail{cuneyt.akcora@umanitoba.ca}

\fundinginfo{NSF DMS 1925346, NSF ECCS 2039716, RGPIN-2020-05665}

\runningauthor{Akcora et al.}

\begin{document}

\maketitle

\begin{abstract}
Blockchain is an emerging technology that has enabled many applications, from cryptocurrencies to digital asset management and supply chains. Due to this surge of popularity, analyzing the data stored on blockchains poses a new critical challenge in data science. 

To assist data scientists in various analytic tasks on a blockchain, in this tutorial, we provide a systematic and comprehensive overview of the fundamental elements of blockchain network models.  We discuss how we can abstract blockchain data as various types of networks and further use such associated network abstractions to reap important insights on blockchains' structure, organization, and functionality.  

\end{abstract}

\section{Introduction}
On October 31, 2008, an unknown person called Satoshi Nakamoto posted a white paper titled “Bitcoin: A Peer-to-Peer Electronic Cash System” to the Cyberpunks mailing list. In eight pages~\cite{nakamoto2008bitcoin}, Satoshi explained the network, transactions, incentives, and other building blocks of a digital currency that he called Bitcoin.

Bitcoin solves the problem of sending and receiving digital currency on the world wide web. \green{However,} the idea of a digital currency is as old as the Internet. For similar purposes, traditional banks and Internet companies have created online payment services, such as Paypal, Visa, and Master. However, a trusted entity intermediates currency flows in these solutions and updates user balances as transactions are processed. Blockchain removes the trusted entity and provides a framework to process transactions and maintain user balances correctly and securely.

Bitcoin and blockchain have been used interchangeably in the past, but Bitcoin is just one financial application among many other use cases of blockchain technology.

Blockchain stores a limited number of transactions (i.e., coin transfers) in a data structure called a block, which is in turn stored in a public ledger. We may consider the ledger as a notebook that has information to calculate user balances written in it. Blockchain represents a user with a blockchain address as a fixed-length string of characters (e.g., 1aw345….). A transaction can be as simple as sending bitcoins from one address to another, along with the required digital signatures. Transaction size increases if more addresses are involved. Blocks contain executed transactions. Block sizes (which determine the number of transactions in a block) are usually small (e.g., 1MB in Bitcoin that allows approximately 3000 transactions).

Blockchain uses an underlying peer-to-peer network to transmit blocks and proposed transactions between blockchain users worldwide. We will elaborate on “users” in later sections; however, users of the first blockchain, Bitcoin, are ordinary web users who can join the network by downloading and installing an application called a wallet.

A copy of the ledger is stored locally at every participant (i.e., node) of the peer-to-peer Bitcoin network. Every user is supposed to check its blockchain copy to learn about user balances. Bitcoin ledger is extended by appending a new block to the end of the chain every 10 minutes through a process that is called mining.
The mining is where transaction approval/verification and coin creation occur, and blockchain removes financial institutions from this process and allows ordinary users to serve in their role. Mining is achieved by solving a cryptographic puzzle by trial and error. A valid solution is a proof that the miner has completed some work and effort to find the answer to the puzzle. The answer itself is an integer called nonce that once appended to the end of block content, the hash of the content satisfies a predetermined difficulty goal. The nonce and the resulting hash are known as proof-of-work, which has its roots in a research article~\cite{dwork1992pricing}.

Mining of each block is an open competition worldwide. Any user may compete to solve the puzzle to earn a handsome sum called the block reward, creating new bitcoins. This process is similar to fiat money printing; however, coins are created in predefined quantities (50 BTC in 2009, the amount halves every four years) at each block and given to the miner. Mining is, by purpose, designed to be difficult so that miners will not litter the blockchain with blocks. However, the difficulty is adaptive and depends on the number of miners and their computing power.
Furthermore, Bitcoin uses an adaptive difficulty level to reduce the frequency of block mining. The blockchain peer-to-peer network participants will receive the latest block in time and be aware of its transactions. Awareness prevents malicious users from spending the same coin multiple times. For example, Bitcoin updates the mining difficulty to create a ten-minute gap between two blocks on average, and Ethereum aims at 12-15 seconds per block.

\begin{figure}
    \centering
    \includegraphics[width=\linewidth]{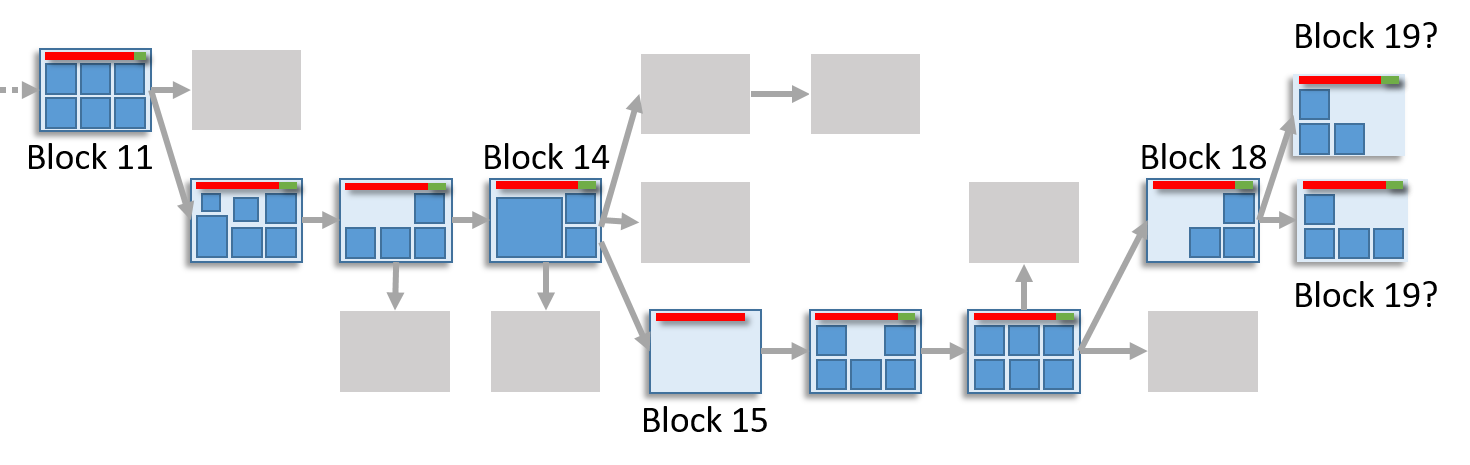}
    \caption{Block structure. We show transactions with squares within blocks. The canonical blockchain is the main chain that contains blocks and continues to grow in time. New blocks are expected to be mined on the top (rightmost part) of this blockchain. Non-transactional block data, such as block hash or nonce, are not shown in this figure, but edges indicate parents. Blocks that we depict in gray are called stale blocks because they do not appear in the canonical blockchain. At the right end of the blockchain, two blocks are competing to be the $19$th block. Both blocks are valid, but eventually, the canonical blockchain will include only one of them, and the other will be a stale block. The long transactions at the top of each block are coinbase transactions (rewards of miners) which contain the sum of newly minted coins (in red) and transaction fees (in green). Block 15 has no transaction other than the coinbase one.}
    \label{fig:blockchain}
\end{figure}

As nodes compete worldwide, miners create candidate blocks and append them to the chain. The competition is resolved by adopting the candidate that appears on the longest blockchain (see Figure~\ref{fig:blockchain}). As miners append more blocks to the longest chain, the earlier blocks are deemed less likely to change, i.e., data in earlier blocks are considered final. 
 
In Figure~\ref{fig:blockchain} a blockchain has 19 blocks (the latest block, $b_{19}$ has two competing candidates).  The longest chain is called the canonical path, and blocks on the chain are deemed valid (i.e., $b_{11}$ to $b_{19}$). Gray boxes show stale blocks that have failed to be a part of the blockchain. An arrow shows a parent-child relationship between blocks. For example, $b_{14}$ is the parent of $b_{15}$ ($b_{15}$ is mined later than $b_{14}$).

Blockchains proliferate, and the ecosystem is in evolution. A diverse set of blockchains have proposed novel mechanisms in storing data and representing users, coin transfers, smart-contract operations, and more. As a result, mining blockchain data requires domain expertise which may seem daunting to data scientists.

In this survey, we take a holistic view and define salient characteristics of six significant blockchains in terms of their data structures: Bitcoin, ZCash, Monero,  Ethereum, Ripple, and IOTA. As Bitcoin is the first blockchain, we first teach the UTXO blockchains and design choices of Nakamoto that have significantly affected how new blockchains function and store data. The Ethereum section covers the basics of smart contracts, tokens, and other decentralized finance constructs. In the Ripple and IOTA sections, we teach their use cases beyond cryptocurrency and platform aspects. We conclude the survey with an overview of graph types found in these blockchains.pears on the longest blockchain

\section{A Taxonomy of Blockchains and Transaction Networks}
After Bitcoin's popularity, various improvements to blockchain have been suggested and implemented in other digital currencies. Bitcoin, so far, has been very conservative in changes to its core technology. A few common points stand out between all blockchain implementations in mining, coding capabilities, and data scalability solutions. 

We can broadly classify blockchains as i) private or public blockchains and ii) currency or platform blockchains. A third discussion point is about first vs. second layer technologies that determine how much data blockchains should store on-chain.

\noindent\textbf{Private, Consortium vs. Public.} Most blockchains, such as Bitcoin and Ethereum, are permissionless. They store data publicly, and any user can mine a block (Ethereum 2.0, when implemented, will change the mining algorithm). Anyone can join the Blockchain network and download blocks to view the transactions in them. When used by companies, a public blockchain would reveal sensitive corporate data to outsiders. Instead, private blockchains, such as Hyperledger fabric by IBM~\cite{cachin2016architecture}, have been developed to restrict block mining rights or data access to one participant (i.e., private blockchain) or a few verified participants (i.e., consortium blockchain).

\noindent\textbf{Currency vs Platform.} Bitcoin and other cryptocurrencies store coin transactions in blocks as data. However, blockchain is oblivious to the type of stored data, which can be multimedia files, weblogs, digitized books, and any other data that we can fingerprint by using a hash function. A hash function takes an unlimited length of data and creates a fixed length (e.g., 256 bits in SHA-256) representation using mathematical operations.

In 1997, Nick Szabo, an American computer scientist, had envisioned embedding what he called smart contracts as \textquote{contractual clauses in the hardware and software to make a breach of contract expensive}~\cite{szabo1997idea}. Almost 20 years later, Blockchain enthusiasts saw a path to implement this vision. Blockchain platforms Neo (2014), Nem (2015), Ethereum (2015), and Waves (2016) have stored and run software code, called smart contracts, on a blockchain. Smart contracts are written in Turing-complete language. However, the block gas limit restricts what transactions can include in the code. For example, on Ethereum, a smart contract cannot be arbitrarily long to encode an algorithm. As a result, blockchain opponents argue that the Ethereum gas limit prevents Turing-complete smart contracts.

The benefits of storing code on a blockchain are multi-fold. Every blockchain network node can read smart contracts, and transactions can execute contracts by sending call messages to them, and all blockchain nodes will run the contract with the message. These aspects mean that smart contracts allow unstoppable, unmodifiable, and publicly verifiable code execution as transactions between blockchain participants.
Any blockchain classification can be flexed to allow public-private functionalities.  For example, Hyperledger Fabric uses smart contracts in permissioned settings. The Ripple credit network is a permissioned blockchain that restricts block mining but stores data publicly.

\noindent\textbf{First vs. Second Layer Technologies.} Over time, blockchains started to run into scalability issues due to limited block sizes and increased user participation. Blockchains developed initial solutions, such as SegregatedWitness,\footnote{https://en.bitcoin.it/wiki/Segregated\_Witness} to leave some of the encryption signatures and other non-transactional data out of blocks to make space for more transactions. Scalability efforts have culminated in second layer solutions, such as the Lightning Network~\cite{poon2016bitcoin}, where the network executes most of the transactions off-the-blockchain. The first layer (i.e., the blockchain itself) only stores a summary of transactions that occur on the second layer.  Second layer networks, such as the Lightning Network, are a hot research area in network analysis that can be the topic of an extensive review article on its own. Due to space limitations, this manuscript will only teach blockchain networks that we can extract from the first layer, i.e., the blockchain itself.

Blockchains have come to contain various data such as sensor messages on IOTA, software code on Ethereum, and international banking transactions on Ripple. We give an outline by considering two major blockchain types: cryptocurrencies and platforms. 
Cryptocurrencies are designed to store financial transactions between addresses. A Blockchain platform stores financial transactions as well but furthermore stores software code in~\textit{smart contracts}.

Blockchains can further be categorized into two broad categories in terms of transaction type: \textbf{account-based} (e.g., Ethereum) and \textbf{unspent transaction output (UTXO)-based} (e.g., Bitcoin, Litecoin) blockchains. Traditionally cryptocurrencies are UTXO-based, whereas platforms are account-based. The difference between UTXO and account-based models has a profound impact on blockchain networks. In the first type of account-based blockchains, an account (i.e., address) can spend a fraction of its coins and keep the remaining balance. An analogy to account-based blockchains is a bank account that makes payments and keeps the remaining balance in the account. A transaction has exactly one input and one output address (an address is a unique identifier on the blockchain transaction network). We may use an address to receive and send coins multiple times. The resulting network is similar to traditional social networks, which implies that we can apply social network analysis tools directly to account networks. In turn, the second type (i.e., UTXO-based blockchains), such as Bitcoin, are the earliest and most valuable blockchains. Bitcoin constitutes around 45-60\% of total cryptocurrency market capitalization. Litecoin has approximately 2\% capitalization. On UTXO based blockchains, nodes (i.e., addresses) are mostly one-time-use only, complicating network analysis. See Section~\ref{sec:limitations} for issues that complicate network analyses.

\begin{remark}
A blockchain platform uses smart contracts to implement complex transactions, resulting in a diverse set of networks.  Traditionally, blockchain platforms have followed an account-based blockchain model. For this reason, we use the terms account blockchain and blockchain platform interchangeably. However, in theory, a platform can also employ a UTXO transaction model. Similarly, a cryptocurrency could use an account-based transaction model. The reader must discern the distinction between account and UTXO based blockchain networks.
\end{remark}

 \begin{table}[!ht]
 \caption{Blockchain types. The private/public columns indicate block mining permissions. IOTA and Ripple are maintained by consortiums that own the mining rights. Many projects (such as Ripple and IOTA) developed smart contract functionality many years after their launch. Although IOTA uses a directed acyclic graph instead of a canonical blockchain, the IOTA transaction model is UTXO based. Privacy coins Monero and Zcash use UTXO models with cryptographic security.}
    \label{tab:blockchainTypes}
    \centering
    \begin{tabular}{c c c c c c c c}
      
         &UTXO&Account&DAG&Platform&Cryptocurrency&Private&Public \\
       \midrule
         Bitcoin&\checkmark&&&&\checkmark&&\checkmark\\
          %Litecoin&\checkmark&&&&\checkmark&&\checkmark\\
        \rowcolor{Gray} ZCash& \checkmark&&&&\checkmark&&\checkmark\\
         Monero&\checkmark &&&&\checkmark&&\checkmark\\
         \rowcolor{Gray}  Ripple& &\checkmark&&\checkmark&&\checkmark&\\
           Ethereum&&\checkmark&&\checkmark&&&\checkmark\\
        \rowcolor{Gray} IOTA& \checkmark&&\checkmark&\checkmark&&\checkmark&\\
         \bottomrule
    \end{tabular}
\end{table}
 
Table~\ref{tab:blockchainTypes} shows the focus of this manuscript and lists the blockchains that we will cover. Specifically, we teach two projects that differ from cryptocurrencies and platforms in critical aspects: Ripple and IOTA. Ripple is a credit network that predates blockchain fame. Ripple keeps a ledger of transactions where participants can trade user-issued currencies along with the native cryptocurrency of the Ripple, which is called XRP. The Ripple network moves currencies across the world, and financial institutions mainly use it.  Recently, Ripple has also implemented smart contracts (called Hooks). We include Ripple in this manuscript since Ripple uses key blockchain technologies, such as hash-based identities and smart contracts, although not a blockchain by definition. 
IOTA Tangle is a directed acyclic graph of transactions. However, the underlying transaction model is UTXO based. IOTA relegates mining to each transaction's creator and does not incur a transaction fee. Although IOTA has its associated cryptocurrency, mainly Internet-of-Things devices use IOTA to share and store data in industrial applications. 

\section{Bitcoin, Monero and ZCash: UTXO Networks}
\label{sec:utxotransaction}
A cryptocurrency transaction is a construct that consumes one or more outputs of previous transactions and creates one or more outputs. As Bitcoin is a good representative of cryptocurrencies, we will use its data to teach this section. Note that Litecoin data is in the same format as Bitcoin, and in the past, we could have parsed them by using the same software library (e.g., Bitcoin4J). We will conclude the section with Monero and Zcash networks.

\begin{figure}
  \begin{center}
    \includegraphics[width=0.38\textwidth]{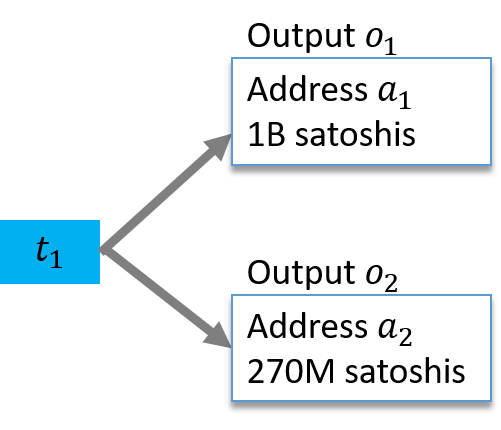}
  \end{center}
  \caption{A coinbase transaction $t_1$ with two outputs. Note that $t_1$ has no input address. Compare this transaction with the ordinary spending transaction of Figure~\ref{fig:spending}. }
  \label{fig:coinbaseRaw}
\end{figure}

Bitcoin stores sequential block data in blk*.dat files on the disk. For example, block1 is serialized and appended into the blk00000.dat file. Next, Bitcoin appends a magic byte to separate block 1 from the upcoming block 2 in the dat file. For example, the Bitcoin mainnet uses \shade{0xD9B4BEF9} as the magic byte. As the number and size of transactions in a block determine the block size, each dat file may contain a different number of blocks.

Each block contains one coinbase transaction and zero or more spending transactions. A \textit{coinbase transaction} is the first transaction of the block and contains the block reward plus a sum of fees from the block's transactions, if any. Note that a transaction may leave nothing as a transaction fee but still get mined in the block. The Bitcoin protocol sets the mining reward. If a miner records the block reward higher than the set amount, the network will reject the block. However, if the miner sets the block reward less than the set amount, the block will be accepted, and the miner will have lost bitcoins. In reality, these bitcoins are lost to the protocol, and the total bitcoin supply decreases from 21 million bitcoins. 

 \begin{figure}
  \begin{center}
    \includegraphics[width=0.46\textwidth]{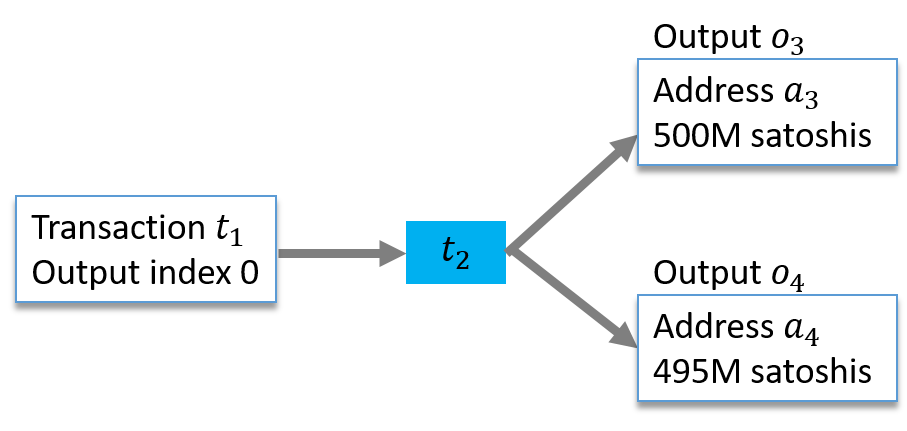}
  \end{center}
  \caption{A spending transaction that consumes a previous output and creates two new outputs.}
  \label{fig:spending}
 \end{figure}

A coinbase transaction has no input but one or more outputs (Figure~\ref{fig:coinbaseRaw}). In a sense, in a coinbase transaction, the miner can write itself a check in the amount of mining reward plus transaction fees. A coinbase transaction increases the total bitcoin supply by the mining reward amount. This is where new coins are created in Bitcoin. For this reason, a coinbase transaction lists no input. All other transactions must list one or more inputs that are being spent. Usually, a coinbase transaction has a single output address, and the address belongs to the miner. Some mining pools pay their miners in coinbase transactions. Such transactions may have hundreds of output addresses where each address belongs to an individual pool member. In Figure~\ref{fig:coinbaseRaw} the coinbase transaction has two outputs that total 1.270B satoshis, out of which 1.25B (12.5 bitcoins) are the mining reward, whereas 20M are the sum of transaction fees. Satoshi is the subunit of bitcoin (as Cent is the subunit of US dollar), and one bitcoin contains 100 million satoshis. As Figure~\ref{fig:coinbaseRaw} shows, an output has two useful information: amount and address. An address can next use the received output of $o_1$ in a spending transaction. Figure~\ref{fig:spending} shows the associated spending transaction that consumes $o_1$ and creates two new outputs. 

A look at Figure~\ref{fig:spending} reveals an interesting point about how Bitcoin records transactions. Transaction $t_2$ gives the consumed output with its previous transaction id (the hash of $t_1$, which created this output) and the output index. From this information, we can deduce that we are spending the zeroth output of $t_1$, but this does not tell us the amount or address ($a_1$). By only linking the previous transaction $t_1$, we can learn the corresponding amount and address. 

From Figure~\ref{fig:coinbaseRaw} we see that the zeroth output of $t_1$ holds 1B satoshis. We will denote the satoshi amount in an output with $A()$, i.e., $A(o_1)=1B$. Next, in Figure~\ref{fig:spending} 500M and 495M satoshis are sent to $a_3$ and $a_4$. The difference between $A(o_1)-[A(o_3)+A(o_4)]=5M$ satoshis are implicitly left (by the transaction creator) as the transaction fee. The miner's job is to validate that output amounts are greater than or equal to the input amounts. Furthermore, the miner must observe all previous blocks and ensure that another transaction has not already spent $o_1$. If the miner makes a mistake and includes an invalid (e.g., already spent output, too great an output amount) transaction, the network will reject the miner's block.

\begin{figure}
    \centering
    \includegraphics[width=\linewidth]{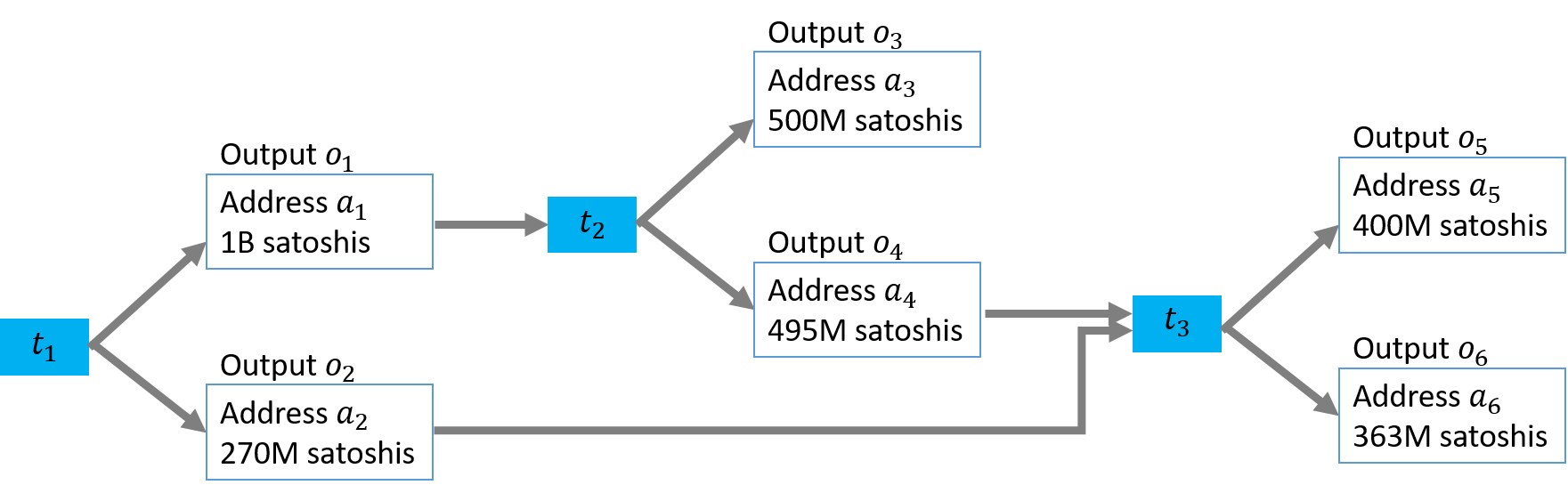}
    \caption{Bitcoin transaction network where we link transaction inputs to previous transaction outputs.}
    \label{fig:outputGraph}
\end{figure}

In practice, starting from an output, it is possible to go back in time and trace the lineage of a coin to reach a coinbase transaction. For example, the output linked graph in Figure~\ref{fig:outputGraph} shows the third transaction $t_3$, which has two inputs. The output $o_6$ which has two lineages: $o_1 \rightarrow o_4 \rightarrow o_6$ and $o_2 \rightarrow o_6$. However, the lineage can be obscured by mixing schemes~\cite{ruffing2014coinshuffle}.

 As well as multiple outputs, a Bitcoin transaction allows multiple inputs. The owner of each input must sign its portion of the transaction to prove an authorized coin transfer. A many-to-many transaction is an interesting data type that we do not encounter in banking and finance; multiple addresses merge their coins and send them to multiple receiving addresses. For the output, a Bitcoin address can be created for free and shared with the sender. Output addresses of a transaction need not belong to the same real-life entity: $a_5$ and $a_6$ may belong to people who have never met in real life. 

The same reasoning is less likely to apply to input addresses.  There can be multiple scenarios considering the inputs. 

\begin{itemize}
    \item The user behind $a_2$ signs the input of $t_3$ and sends the partially signed transaction to the user behind $a_4$, who signs its input and finally sends the transaction to the mempool. This scenario implies that the users are different people but know and communicate with each other.
    \item The user behind $a_2$ signs the input of $t_3$ and sends the partially signed transaction to a public forum where other users can add their inputs, and the transaction is finally sent to the mempool by one of them. This scenario implies that the users are different people, do not know but communicate with each other.
    \item A single user owns both $a_2$ and $a_4$. The user signs both inputs and sends the transaction to the mempool.
\end{itemize}

The tree cases mean that we can never be sure about the joint ownership of inputs. However, in practice, most of the time, all input addresses belong to the same user. We can use this information to link addresses across transactions. 

\begin{remark}
When a transaction sends coins to an address, the address appears in the transaction, but the public key of the receiver address is unknown. The receiver discloses their public key only when spending the received coins, and any user in the network can hash the key to verify that the hash equals the address. The matching hash is additional proof that the assets belong to the receiver. In a P2PSH (pay to script hash) address (which starts with '3'), the owner must show the script whose hash equals the address.
\end{remark}

\subsection{Graph Rules for UTXO Blockchains}
\label{sec:graphrules}
\begin{figure}
    \centering
    \includegraphics[width=0.7\linewidth]{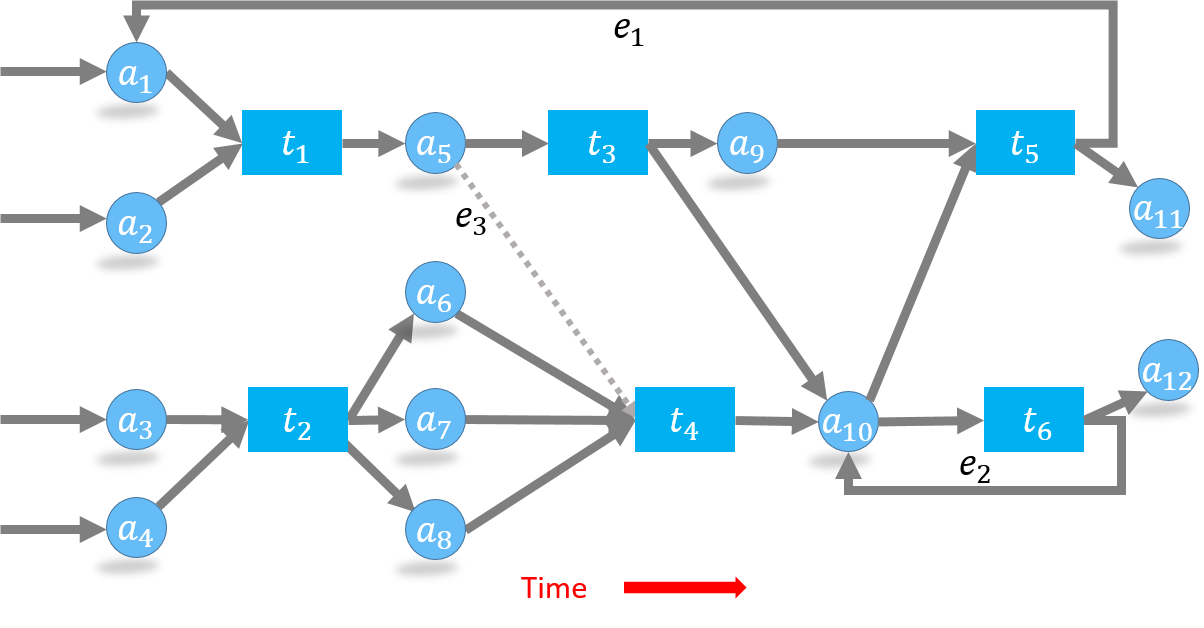}
    \caption{A blockchain network with six transactions and 12 addresses. Addresses $a_1,a_2,a_3,a_4$ receive coins from earlier transactions that are not shown here.  Transaction $t_5$ uses a previously used address $a_1$ in edge $e_1$. Edge $e_2$ denotes an address reuse where the change is sent back to the sender address $a_{10}$. Dashed edge $e_3$ \green{cannot be created} with the given transactions, because $a_5$ receives only a single output in $t_1$. In this network, outputs of $t_5$ (to addresses $a_1$ and $a_{11}$) and $t_6$ (to addresses $a_{12}$ and $a_{10}$) remain unspent.}
    \label{fig:blockchaingraph}
\end{figure}

Before modeling the Bitcoin network, we emphasize \textbf{three graph rules} that restrict how transactions can transfer coins. These rules are due to Bitcoin design choices made by Satoshi Nakamoto. For clarity, we will replace outputs with the associated addresses and work with the toy graph shown in Figure~\ref{fig:blockchaingraph}.

\begin{enumerate}
\item {\bf Balance Rule}:  Bitcoin transactions consume and create outputs. Outputs are indivisible units; we can spend each output in one transaction only. Consequently, we must spend coins received from one output in a single transaction. Any amount that we do not send to an output address is considered a transaction fee, and the miner collects the change as the transaction fee. To spend a portion of the coins in output and keep the change, we must create a new address and send the remaining balance to this new address. Another option is to use our existing address as one of the output addresses and re-direct the balance (as done in edge $e_2$ in Figure~\ref{fig:blockchaingraph}). As a community practice, reuse of the spender's address (i.e., \textit{address reuse}) is discouraged. As a result, most address nodes appear in the graph two times: first when they receive coins and second when they spend them.
\item {\bf Source Rule}: We can merge input coins from multiple transactions and spend them in a single transaction (e.g., the address $a_{10}$ receives coins from $t_3$ and $t_4$ to spent in $t_5$ and $t_6$ in Fig.~\ref{fig:blockchaingraph}). However, $a_{10}$ could have spent all its coins in a single transaction as well.

\item {\bf Mapping Rule}:\label{rule:2} In a transaction the input-output address mappings are not explicitly recorded. For instance, consider the transaction $t_2$ in Fig.~\ref{fig:blockchaingraph}. The output to address $a_6$ may come from either $a_3$ or $a_4$. We can make an analogy with lakes where in-flowing rivers (inputs) bring water (coins) to the lake (transaction), and outgoing emissaries (outputs) take the water (coin) out.
\end{enumerate}

Besides the three rules, in our research, we have encountered the following Bitcoin spending practices:

\begin{itemize}
    \item A transaction may list the same address in multiple outputs. We have encountered transactions where all outputs have the same address. As the address must be recorded more than once, this (spam) practice unnecessarily increases the transaction size.
    \item In time, an address may receive outputs from multiple transactions. Typically, an address collects all outputs and spends them at once. In rare cases, an address can receive outputs again after it has spent its coins. For example, edge $e_1$ in Figure~\ref{fig:blockchaingraph} brings an output to $a_1$ after $a_1$ has spent its coins in $t_1$.
    \item Address reuse, as shown in $e_2$, is widespread. Although the community frowns upon address reuse, users are still reluctant to create a new address for change. Hierarchically deterministic wallets facilitate automated address creation (\url{https://en.bitcoin.it/wiki/Deterministic_wallet}). However, address reuse did not abate.
    \item Community practice is to wait for six confirmations (blocks) after receiving a transaction output before spending it. The practice protects bitcoin buyers against history reversion attacks. Thereby, we must spend coins of an output in a minimum of six blocks after they are received ($6\times 10=60$ minutes later). A day sees 144 blocks; a coin should move at most 24 times \green{(in 24 blocks)}. In reality, however, once a transaction reaches the mempool, it can be considered final, and absent a double-spending, its coins can be used in another transaction immediately, even in the same block. 
\item It is possible to have $n$ ordered transactions in a single block, where $1< x< y< n$ and $t_y$ consumes the outputs generated by $t_x$. As a result, some coins move in the Blockchain network more than 144 times a day.
\end{itemize}

\begin{remark}
In theory, Monero, Dash, and ZCash privacy coins are  UTXO blockchains; their Blockchain networks are similar to Bitcoin's. In practice, privacy coins employ cryptographic techniques to hide node and edge attributes in the blockchain network. For example, ZCash hides information in its shielded pool, whereas Monero adds decoy UTXOs to the input UTXO set. This section explains how to model the blockchain network when the node and edge attributes are public (as in Bitcoin). 
\end{remark}

Data scientists model Bitcoin transaction networks as address or transaction graphs. Address graph omits transactions, and the transaction graph omits addresses. Both approaches are influenced by traditional social network analysis, which employs graphs with one node type only.  Additionally, one can extract \textit{chainlet} substructures from the blockchain network. In the following sections, we will cover these graph models.

\subsection{UTXO Transaction graph}
\begin{figure}
  \begin{center}
    \includegraphics[width=0.38\textwidth]{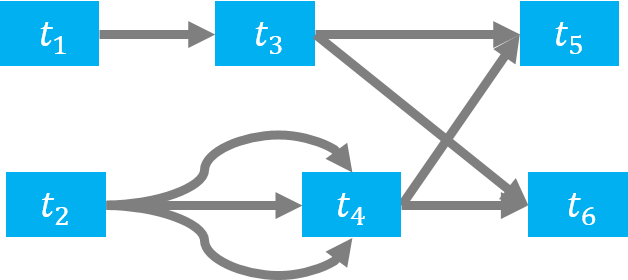}
  \end{center}
  \caption{Transaction graph representation of the Blockchain network in Figure~\ref{fig:blockchaingraph}.}
   \label{fig:transactionGraph}
\end{figure}

Transaction graphs omit address nodes from the transaction network and create edges among transactions only.  Figure~\ref{fig:transactionGraph} shows the transaction graph of the network shown in Figure~\ref{fig:blockchaingraph}. The most important aspect of the transaction graph is that a node can appear only once. There will be no future edges that reuse a transaction node.

The transaction graph contains far fewer nodes than the network it models. We can immediately observe a few drawbacks from Figure~\ref{fig:transactionGraph}. By omitting addresses, we lose the information that $t_5$ and $t_1$ are connected by $a_1$. The address reuse of $a_{10}$ is hidden in the transaction graph as well. Additionally, unspent transaction outputs are not visible; we cannot know how many outputs are there in $t_5$ and $t_6$. Similarly, if $t_3$ had an unspent output, we would not learn this information from the graph. In Bitcoin, many outputs stay unspent for years; the transaction graph will ignore all of them. 

The advantages of the transaction graph are multiple. First, we may be more interested in analyzing transactions than addresses. For example, anti-money laundering tools aim at detecting mixing transactions, and once they are found, we can analyze the involved addresses next. Many chain analysis companies focus their efforts on identifying e-crime transactions. Second, the graph order (node count) and size (edge count) are smaller, which is better for large-scale network analysis. The reduction is because, on UTXO networks, transaction nodes are typically less than half the number of address nodes. For example, Bitcoin contains 400K-800K unique daily addresses but 200K-400K transactions only. As we will explain in the next section, the address graph contains many more edges than the transaction graph. 

\subsection{UTXO Address Graph}

The address graph is the most commonly used graph model for UTXO networks. The address graph omits transactions and creates edges between addresses only. Address nodes may appear multiple times, which implies that addresses may create new transactions or receive coins from new transactions in the future. 

\begin{figure}
  \begin{center}
    \includegraphics[width=0.48\textwidth]{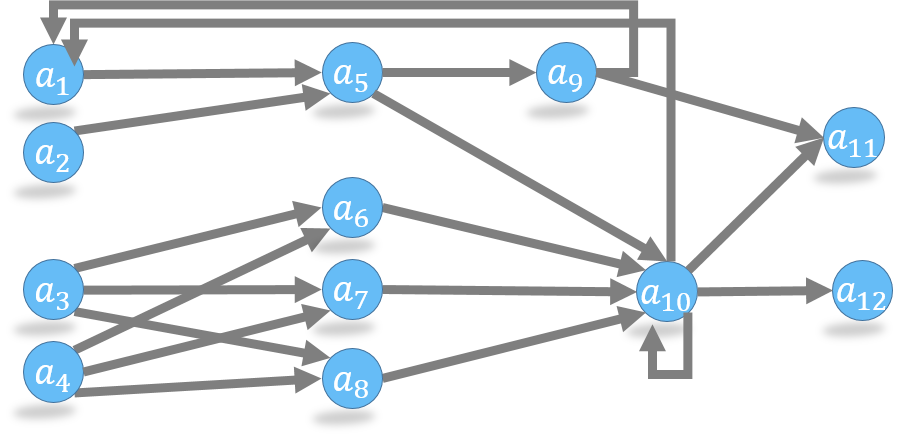}
  \end{center}
  \caption{Address graph representation of the Blockchain network in Figure~\ref{fig:blockchaingraph}.}
   \label{fig:addressgraph}
   \end{figure}

Address graphs are larger than transaction graphs in node and edge counts. As the mapping rule states, a UTXO transaction does not explicitly create an edge between input and output addresses in the blockchain transaction network. When omitting the intermediate transaction node, we cannot know how to connect input-output address pairs. As a result, data scientists create an edge between every pair.  For example, in Figure~\ref{fig:addressgraph} we create six edges between inputs ($a_3,~a_4$) and outputs ($a_6,~a_7,~a_8$). If there are few addresses in the transaction, this may not be a big problem. However, large transactions can easily end up creating millions of edges. For example, the highest number of inputs in a Bitcoin transaction was 20000 (821 on Litecoin), whereas the highest number of outputs in Bitcoin was 13107 (5094 on Litecoin). The address graph approach will have to create one million edges for a transaction with one thousand inputs and one thousand outputs.

Graph size is not the only problem. The address graph loses the association of input or output addresses. For example, the address graph in Figure~\ref{fig:addressgraph} loses the information that edges $a_3$ and $a_4$ were used in a single transaction; address graph edges would be identical if the addresses had used two separate transactions to transfer coins to $a_6~a_7$ and $a_8$. We can solve this issue by adding an attribute (e.g., transaction id) to the edge; however, this requires additional edge features. 

Address graph edge weighting is done as follows. Consider a transaction $t$ with its input addresses $I$ and output addresses $O$. We will denote the amount an address $a_x$ sends to or receives from $t$ as $w_{x}$. Amount of coins in an address is denoted as $A(a_x)$. The edge between an input address $a_i\in I$ and an output address $a_j\in O$ is assigned an edge weight as $$w_{i\rightarrow j}=A(a_i)\times \dfrac{A(a_j)}{\sum_{a_x\in O}{A(a_x)}}.$$

\begin{example}
Consider a transaction $t$ with input addresses $I=\{a_1,a_2\}$ and amounts as $A(a_1)=1$, $A(a_2)=3$ bitcoin. The transaction has two output addresses $O=\{a_3,a_4\}$ and their associated amounts as $A(a_3)=0.9$, $A(a_4)=2$ bitcoin. The transaction fee is left as the difference between inputs and outputs: $1+3-(0.9+2)$ bitcoin. When creating the address graph, $w_{2\rightarrow 3}$ is weighted as $3 \times (0.9/2.9)$, whereas $w_{2\rightarrow 4}=3 \times (2/2.9)$.
\end{example}

Compared to the transaction graph, the address graph loses less information from the Blockchain network. For example,  the address graph does not fail to record past address reuse (edge from $a_9$ to $a_1$) and change address reuse (as self-loop at $a_{10}$). Furthermore, unspent output addresses remain visible in the graph. 

\begin{remark} \textbf{Address reuse:}
A blockchain address can be used in multiple transactions as a coin sender or receiver. However, address reuse is discouraged. Ideally, a Bitcoin user should create a new address each time it receives bitcoins in a transaction. Address clustering methods aim at linking multiple addresses to identify real-life entities behind the addresses. 
\end{remark}

Once we create the address graph, we may be inclined to run traditional network science tools and algorithms on it. However, some of these methods are ineffective in blockchain networks. We can count address clustering, motif analysis, and core decomposition among these methods. UTXO networks are sparse and devoid of closed triangles. Furthermore, address reuse is discouraged on UTXO blockchains as a community practice; most addresses appear in two transactions (i.e., receiving and spending coins).  As a result, off-the-shelf Data Science algorithms and software libraries are relatively inefficient on address graphs. 

For example, network motif analysis~\cite{milo2002network} aims at finding repeating subgraphs of specific orders (usually three nodes). Searching for motifs in Figure~\ref{fig:addressgraph} will try to find shapes without considering how specific addresses appear together as inputs or outputs. Addresses $a_3~a_6$, and $a_4$ will most likely never form a triangle. Also, a triangle between $a_6,~a_7$, and $a_8$ is very unlikely. As a second example, motifs do not consider which addresses can be active. For instance, once $a_3$ spends its output in a transaction without receiving a new output, it will never create another edge in the future. Keeping such nodes in memory and searching for their future edges are unnecessary. 

Past address reuse is discouraged, but motifs do not use this information and search for (non-existent) edges to past addresses in a large address graph. As a third example, consider that many coins are moved in the network more than 100 times a day, and each time they are received in newly-created addresses. The resulting graph will be huge, but there will be almost no closed edge triangles. 

Due to these issues, we argue that we should model UTXO networks as a forward branching tree rather than networks. More importantly, Data Scientists should develop tools to incorporate domain practices such as aversion to address-reuse. 

\subsection{Monero and Zcash Transaction Networks}

Monero and Zcash are two UTXO based cryptocurrencies that hide transaction details from the public. For this reason, Monero and Zcash are called privacy coins.

\subsubsection{Monero Networks}
\begin{figure}
    \centering
    \includegraphics[width=0.6\linewidth]{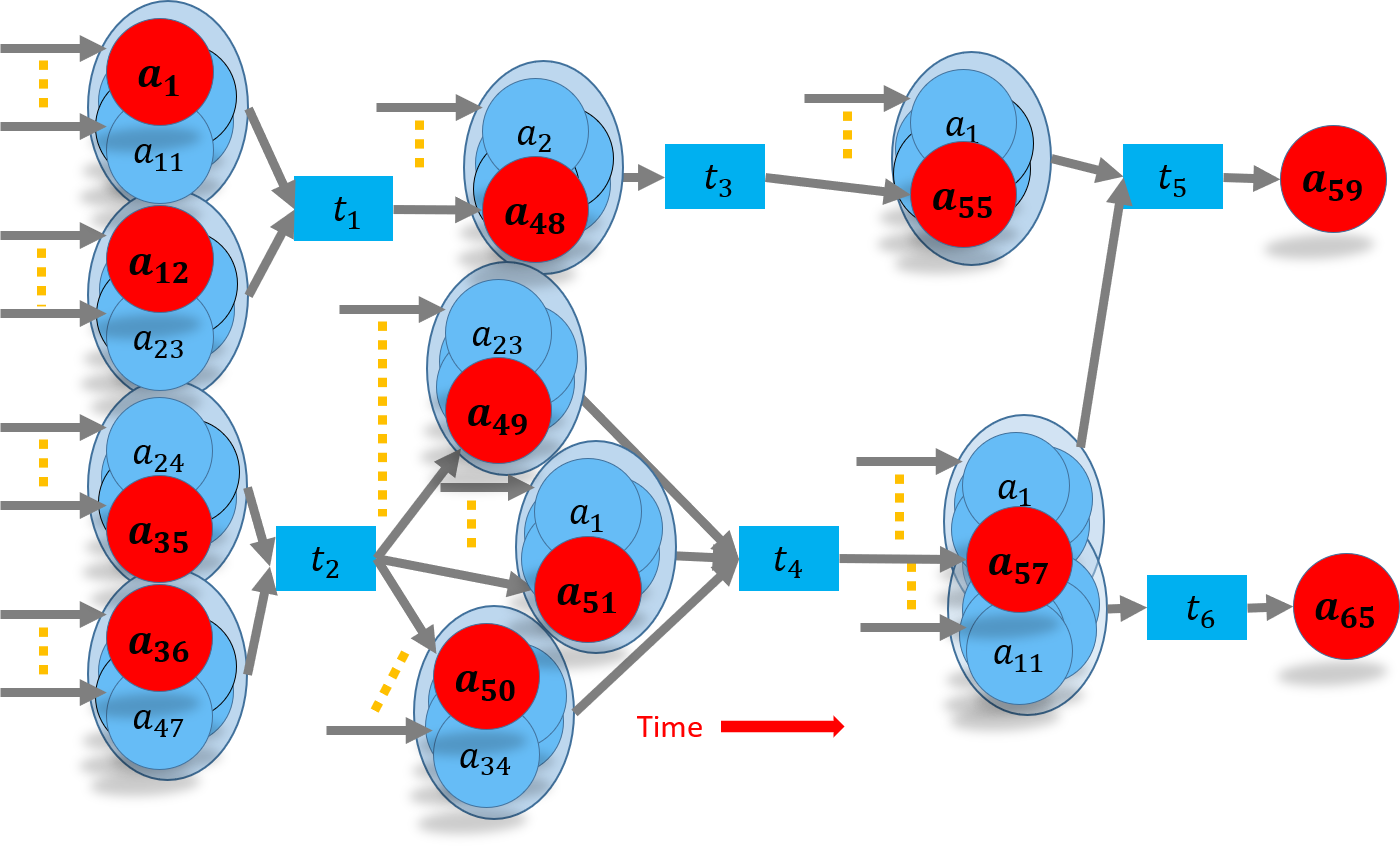}
    \caption{Monero transaction network. We show the actual transaction addresses with red nodes and decoy UTXO addresses with blue nodes. Each ring surrounds nodes in light blue. Dashes indicate the remaining UTXOs that are linked to previous transaction outputs. Note that each input ring has exactly ten decoys and one actual input. However, outputs are listed individually and not in a ring. Address $a_1$ is truly spent in $t_1$, but listed as a ring member in $t_5$ as well. By only looking at transaction inputs, one cannot learn which transaction spends the coins at $a_1$. Since the RingCT update, Monero has hidden the UTXO amount and address as well. The reader should note how difficult, if not impossible, it would be to link addresses to transactions in such a graph.}
    \label{fig:monero}
\end{figure}

Monero (created in 2014) uses ring signatures to hide inputs of a transaction. In a ring signature scheme~\cite{noether2015ring}, any one of the $m$ members of the ring can sign a document. Any third party can publicly verify the signature but cannot deduce which ring participant signed it. Monero treats each transaction input as a document that ring members will sign. The process to create a transaction is carried out as follows: 

\begin{enumerate}
    \item Choose one or more of your UTXOs that you want to spend in a transaction.
    \item For each UTXO that you plan to spend, \textit{choose} 10 foreign UTXOs  (i.e., ring member UTXOs that do not necessarily belong to you) that you will use as decoys.
    \item Prepare the ring signature by using the private key of your address.
    \item After signing each input, forward the transaction to the Peer-to-Peer network.
\end{enumerate}

The reader should note that decoy UTXOs can belong to any other user, and the transaction creator does not need the consent of these users to include their UTXOs as decoys in a ring. Choosing the best decoy UTXOs to hide the identity of the actual input UTXO is an active research area.  

Monero initially allowed using 0-decoys, i.e., not adding any decoys to the ring. In this case, a Monero transaction was equivalent to a Bitcoin transaction where all information is public. Monero also allowed a user to set the ring size, which could be very big. Privacy research has shown the drawbacks of this liberal approach~\cite{kumar2017traceability,moser2018empirical}. First, 0-decoy transactions jeopardize the privacy of other blockchain users~\cite{moser2018empirical} since a 0-decoy transaction input reveals that no other transaction could have spent it. As a result, we can remove the input from the rings of future transactions. Second, we can use a meta-analysis on similar ring sizes and link transactions of the same user. 

Monero banned 0-decoy transactions in 2016, incremented the mandatory ring size from 5 to 7 and eventually to 11. With the RingCT update, Monero has also hidden UTXO addresses and amounts. As a result, very few pieces of transactions are visible to the public, making it an ideal privacy coin. Figure~\ref{fig:monero} shows a Monero address-transaction graph. Note that although inputs list 11 member rings, outputs are listed individually. However, the receiving address and received amount are hidden in the UTXO.  Monero uses commitments and range-proofs to make sure that amounts are spent correctly (See Chapter~4 in  \cite{alonso2018monero}).

\subsubsection{Zcash Networks}
Zcash uses zero-knowledge proofs to hide information about some transactions. In cryptography, a zero-knowledge proof is a method by \textquote{which one party can prove to another party that they know a value $x$, without conveying any information apart from the fact that they know the value $x$}.~\footnote{\url{https://z.cash/technology/}}

We can broadly categorize Zcash transactions into shielded and public transactions. Public transactions are identical to Bitcoin transactions where amounts, addresses, and all other pieces of information are public. Shielded transactions hide all transaction information from the public by using zero-knowledge proofs. Addresses that are used in public transactions are called \textit{t-addresses} and start with the letter \textquote{t} such as \shade{t1K2UQ5VzGHGC1ZPqGJXXSocxtjo5s6peSJ}. Shielded transactions use shielded (also called private) addresses that start with the letter \textquote{z}.  

\begin{figure}
    \centering
    \includegraphics[width=0.6\linewidth]{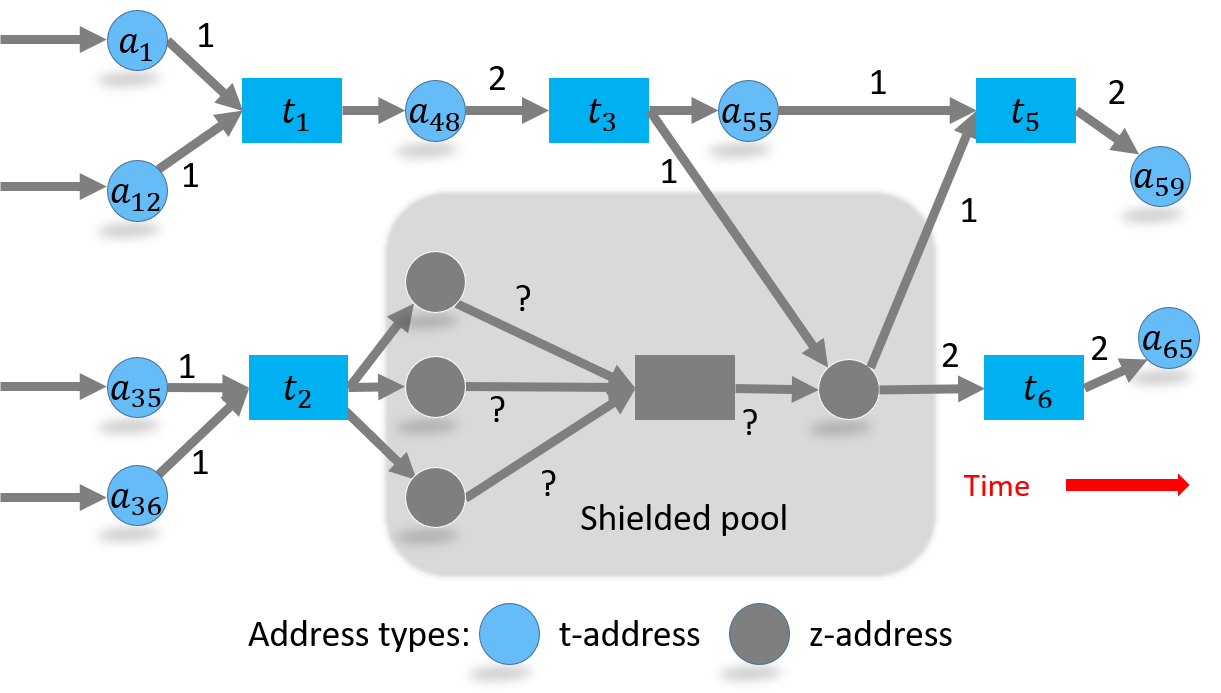}
    \caption{A Zcash address-transaction graph that contains six transactions. Amounts on edges are coins (we assume zero transaction fees). Although we show the shielded pool nodes and edges here, in reality, the pool is not visible to the public at all. Transactions which involve $z$-addresses are computationally costly to create and validate. As a result, only around 10\% of Zcash transactions are shielded~\cite{biryukov2019deanonymization}. }
    \label{fig:zcashnetwork}
\end{figure}

With $t$ and $z$ addresses, there are five Zcash transactions types. We will show these types by using Figure~\ref{fig:zcashnetwork}, and explain them as follows:

\begin{enumerate}
    \item $t$-to-$t$ (public) transactions: all transactions details are public ($t_1$ in Figure~\ref{fig:zcashnetwork}). 
    \item $t$-to-$z$ (shielding) transactions: $t$ address and its sending amount are public, $z$ address is encrypted ($t_2$ in Figure~\ref{fig:zcashnetwork}). 
    \item $z$-to-$t$ (de-shielding) transactions: $t$ address and amount it receives are public, $z$ address is encrypted ($t_6$ in Figure~\ref{fig:zcashnetwork}).
    \item $z$-to-$z$ (private) transactions: the addresses, transaction amount, and the memo field are all encrypted and not publicly visible, i.e., $t$ (without an index) in the shielded pool of Figure~\ref{fig:zcashnetwork}.
    \item $tz$-to-$tz$ (mixed) transactions: $z$-addresses are involved, but there are public inputs or outputs in the transaction ($t_3$ and $t_5$ in Figure~\ref{fig:zcashnetwork}).  
    
\end{enumerate}

A few design choices have made an impact on Zcash transactions. First, the Zcash protocol includes a consensus rule that coinbase rewards must be sent to a shielded address, and typically these coins are forwarded to $t$-addresses very soon.  Second, Zcash initially used a zero-knowledge scheme that supported at most two hidden inputs and two hidden outputs. As a result, including more than two input or output addresses required more cryptographic work, which was costly. A newer scheme called sapling does not have this limitation.

Both Monero and Zcash are much more costly than Bitcoin in terms of computational costs and resource usage. Monero transactions store ten decoys for each actual input, which considerably increases transaction size on disk, and Zcash transaction validation takes too much memory and time. As a result, online exchanges avoid storing balances of $z$-addresses. However, Monero and Zcash both use UTXO models that we can analyze by graph analysis tools developed for Bitcoin.

\subsection{Chainlets for UTXO Transaction Networks}

Subgraph encoding on the Blockchain network is an alternative to address and transaction graph approaches. 

In traditional graphs, we consider nodes and edges the building blocks because nodes are created in time and may establish new edges at various time points. However, UTXo transactions create multiple edges at once. That is, we can think of a transaction itself as a building block of the network. With its nodes and edges, a transaction represents an immutable decision that is encoded as a substructure on the UTXO network. Rather than using individual edges or nodes, we can use this substructure as the building block in network analysis. If we consider multiple transactions and their connections through addresses, we may extend the substructure idea to blockchain subgraphs by considering multiple transactions. We use the term \textit{chainlet} to refer to such subgraphs~\cite{akcora2017chainlet}.   

Consider a UTXO network with transaction and address nodes. The node set $V=\{\text{address},\text{transaction}\}$. An edge $e\in E$ connects an address $a_x$ to a transaction $t_y$ (i.e., $a_x\rightarrow t_y$) or a transaction $t_y$ to an address $a_z$ (i.e., $t_y \rightarrow a_z$). This implies that there are no edges between two nodes of the same type. 

A UTXO chainlet $\mathcal{G'}=(V', E',B)$ is a subgraph of $\mathcal{G}$, if $V'\subseteq V$ and $E'\subseteq E$.  If $\mathcal{G'}=(V', E',B)$ is a subgraph of $\mathcal{G}$ and $E'$ contains all edges $e_{u,v}\in E$ such that $(u,v) \in V'$, then $G'$ is called an induced subgraph of $G$.  
\begin{figure}
\centering
\includegraphics[width=\textwidth]{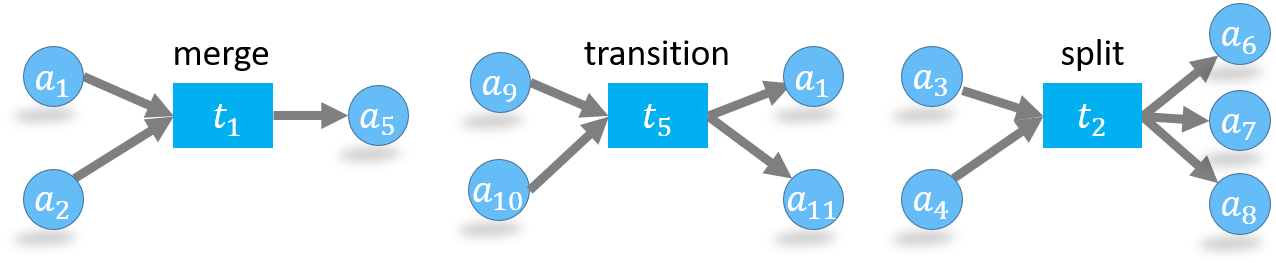}
\caption{Examples of the first order Merge ($\chainlet{2}{1}$), Transition ($\chainlet{2}{2}$) and Split ($\chainlet{2}{3}$) $k=1$ chainlets from the UTXO network of Figure~\ref{fig:blockchaingraph}. An address can appear in multiple chainlets, each time spending or receiving a different UTXO. Transaction nodes can appear in a single chainlet only. 
}
\label{fig:aggregateTypes}
\end{figure}

Let $k$-chainlet $\mathcal{G}_k=(V_k, E_k,B)$ be a subgraph of $\mathcal{G}$ with $k$ nodes of type \enquote{transaction}. If there exists an isomorphism between $\mathcal{G}_k$ and $\mathcal{G}'$, $\mathcal{G}'\in \mathcal{G}$, we say that there exists an occurrence of $\mathcal{G}_k$ in $\mathcal{G}$. A $\mathcal{G}_k$ is called a blockchain \textit{$k$-chainlet}.

As a starting point, we can focus on the first order chainlet ($k=1$), which consists of a single transaction node and the address nodes. The first order chainlet is, by definition, a substructure of the network. We denote a chainlet of $x$ inputs and $y$ outputs with $\chainlet{x}{y}$.  A natural classification of first-order chainlets can be made regarding the number of inputs $x$ and outputs $y$ since there is only one transaction involved. We can quickly identify three main types of first-order chainlets.

If the transaction merges input UTXOs, it will have a higher number of inputs than outputs. We call these \textit{merge} chainlets, i.e., $\chainlet{x}{y}$ such that $x>y$, which show an aggregation of coins into fewer addresses. Two other classes of chainlets are \textit{transition} and \textit{split} chainlets with $x=y$ and $x<y$, respectively, as shown in Figure~\ref{fig:aggregateTypes}. We refer to these three chainlet types as the \textit{aggregate chainlets}.

\begin{figure} 
\centering
\includegraphics[width = 0.9\linewidth]{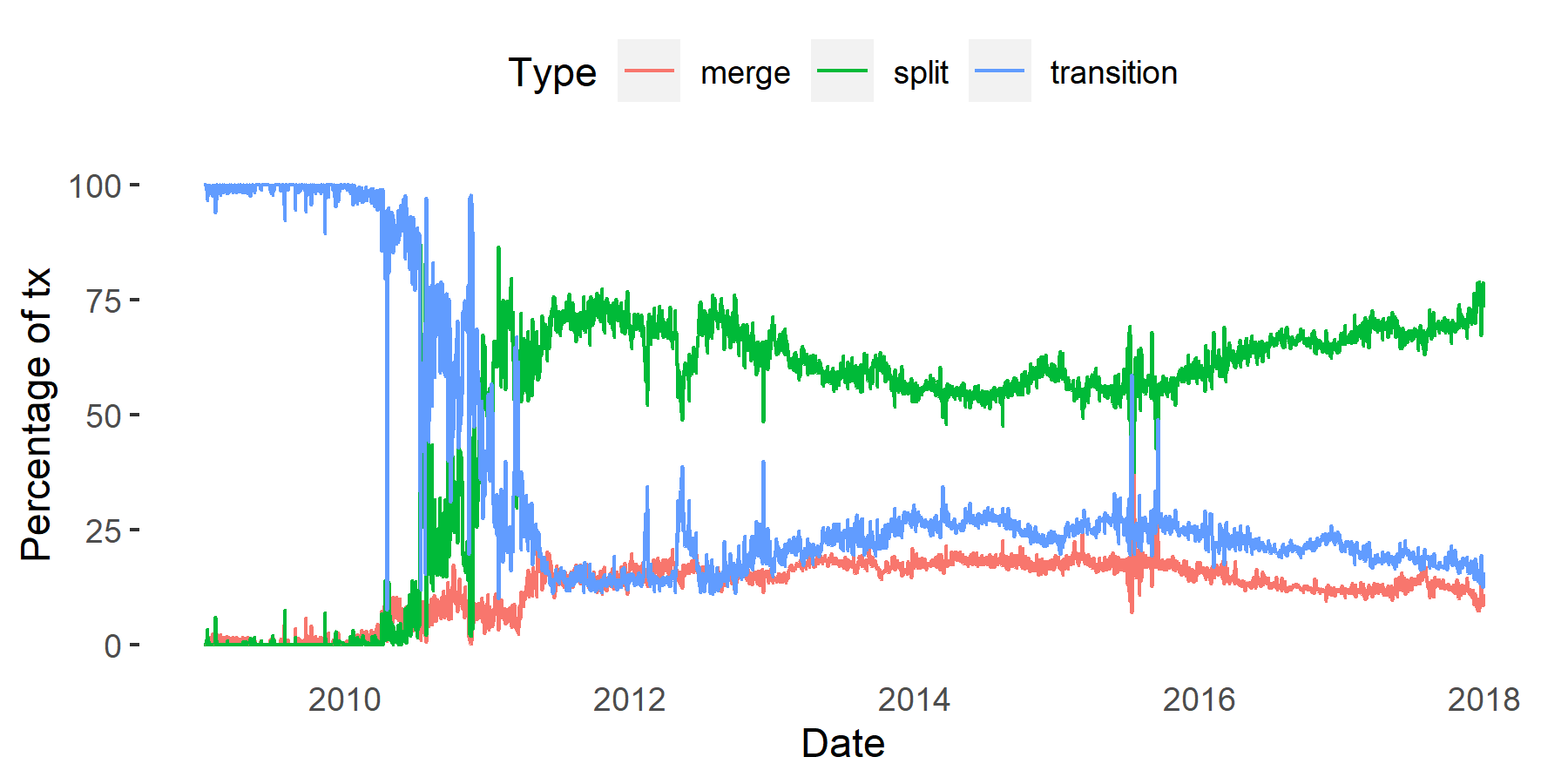}
\caption{Percentage of daily aggregate chainlets in Bitcoin. Splits constitute around 75\% of all transactions.}
\label{fig:chainletsTypesTime}
\end{figure}

Figure~\ref{fig:chainletsTypesTime} visualizes the percentage of aggregate Bitcoin chainlets in time. For example, the transition chainlets are those $\chainlet{x}{x}$ for $x \geq 1$.  Figure~\ref{fig:chainletsTypesTime} shows that the Bitcoin network, starting as an unknown project, stabilized only after summer 2011. From 2014 and on-wards, the split chainlets continued to rise steadily, compared to merge and transition chainlets. Spam attacks on the Bitcoin blockchain, which created too many transactions on the mempool to slow down transaction processing and force a block size increase, were visible around late 2015.

\begin{figure}
    \centering
    \includegraphics[width=\linewidth]{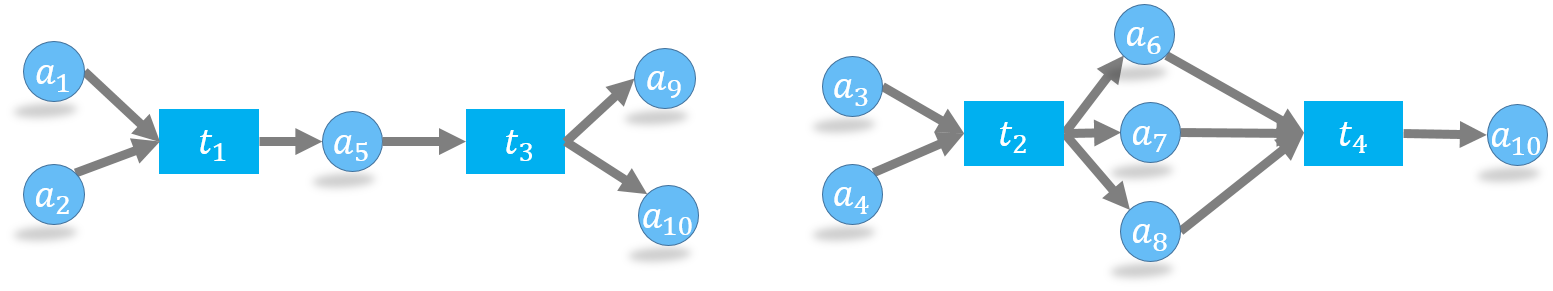}
    \caption{Two second order ($k=2$) chainlets  from the Blockchain network of Figure~\ref{fig:blockchaingraph}.}
    \label{fig:secondorderchainlet}
\end{figure}

 Extending this discussion, higher-order chainlets, as shown in Figure~\ref{fig:secondorderchainlet}, can be classified in terms of their shapes.

\subsection{Occurrence and Amount Information in Chainlets} 

Chainlets provide two lenses to look at the transaction network; we can consider amounts or counts of chainlets.

For a given time granularity, such as one day, we can take snapshots of the blockchain network and extract chainlets from it. From the Blockchain network snapshot for a given granularity (e.g., daily), we mine two information for each chainlet type: \textit{amount} to store volume of coin transfers by using the chainlet and \textit{occurrence} to store instances (i.e., counts) of the chainlet. 

\begin{figure} 
\centering
\includegraphics[width = 0.5\linewidth]{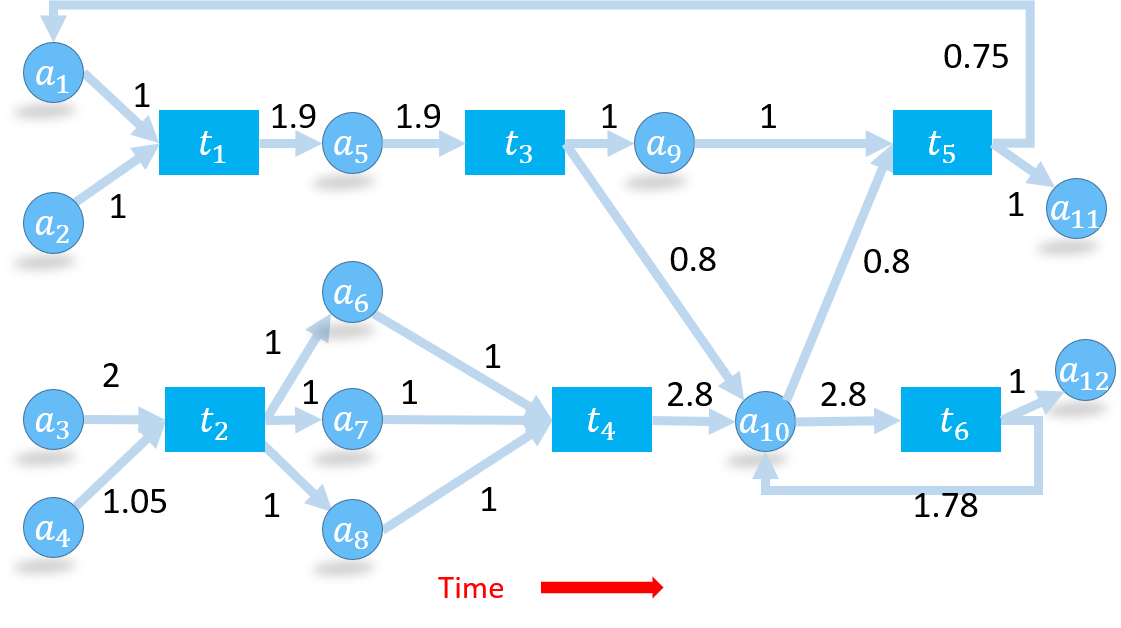}
\caption{A UTXO transaction network with amounts. Transaction fees range from 0.02 to 0.1 coins.}
\label{fig:amountgraph}
\end{figure}

For example, Figure~\ref{fig:amountgraph} contains six transactions and 12 addresses. However, if we look at the first order chainlets, we find five unique chainlets: 
Transaction $t_1$ creates a split chainlet $\chainlet{2}{1}$. Transaction $t_2$ creates split chainlet $\chainlet{2}{3}$. Transactions $t_3$ and $t_6$ create split chainlets $\chainlet{1}{2}$. Transaction $t_4$ creates merge chainlet $\chainlet{3}{1}$. Transaction $t_5$ creates transition chainlet $\chainlet{2}{2}$.

From these chainlets, we can create two $3\times 3$ matrices to hold occurrence and amount information as 
\begin{equation*}
O = 
\begin{bmatrix}
0 & 2 & 0 \\
1 & 1 & 1 \\
1 & 0 & 0
\end{bmatrix},~
A = 
\begin{bmatrix}
0 & 3.58 & 0 \\
1.9 & 1.75 & 3 \\
2.8 & 0 & 0
\end{bmatrix}.
\end{equation*} where $\mathcal{O}[i,j]$ and $\mathcal{A}[i,j]$  store the number and transferred amount of chainlets of the type $\chainlet{i}{j}$, respectively, where $i\geq 1$ and $j\geq 1$. For example, transactions $t_3$ and $t_6$ are stored in $\mathcal{O}[{1,2}]=2$.  The amounts output in these transactions are $\mathcal{A}[1,2]=1+0.8+1+1.78 = 3.58$. 

\begin{remark}
A coinbase transaction has no input address but $\geq 1$ output addresses. If we plan to use the occurrence matrix $\mathcal{O}$ to store coinbase transactions, we need to extend the matrix with $i=0$ as well. Coinbase transactions would then be stored in $\mathcal{O}[{0,j}]$.
\end{remark}

For the transaction network in Figure~\ref{fig:amountgraph}, $i=j=3$ suffices to store the matrices since transactions have at most three input addresses and three output addresses. However, in a real blockchain matrix, dimensions can easily reach thousands. UTXO blockchains restrict the number of input and output addresses in a transaction by limiting the block size (1MB in Bitcoin), but the number of inputs and outputs can still reach thousands. As a result, we can have large  chainlets (e.g., $\chainlet{1000}{200}$, $\chainlet{901}{200}$ or $\chainlet{100}{951}$). Consider the case where we choose to create $1000\times 1000$ to store the occurrence and amount matrices with 1 million cells. Having so many cells is neither useful nor practical since most transaction types will not exist in the network. In turn, corresponding matrix cells in occurrence and amount would hold many zero values, implying that the matrices would be very sparse. As an alternative, we can choose a suitable value for matrix dimensions.

For the optimal matrix dimension $N$, we have analyzed the history of Bitcoin and Litecoin. We have found that \% 91.38 of Bitcoin and \% 91.27 of Litecoin  chainlets have $N$ of 5 (i.e., $\chainlet{x}{y}$ s.t., $x<5$ and $y<5$) in average for daily snapshots. This value reaches \% 98.10 and \% 96.14 for $N$ of 20, for the respective coins. We have selected $N$ of 20 since it can distinguish a sufficiently large number (i.e., 400) of chainlets and still offers a dense matrix. 

The choice of $N$ requires a strategy to deal with transactions whose dimensions are bigger than $N$. We record these chainlets in the last columns/rows of the matrices, which are defined as follows:

\begin{definition}{Amount Matrix}. We denote the total amount of coins transferred by a chainlet $\chainlet{x}{y}$ in a graph snapshot as $\vol{\chainlet{x}{y}}$. Amount of coins transferred by chainlets in the graph snapshot are stored as an $N\times N$-matrix $\mathcal{A}$ such that for $i\leq N,j\leq N$  \[
\mathcal{A}[i,j]=  \left\{
\begin{array}{l l}
\vol{\chainlet{i}{j}} & \quad \text{if $i<N$ and $j<N$, }\\
\sum\limits_{z=N}^{\infty}\vol{\chainlet{i}{z}} & \quad \text{if $i<N$ and $j =N$,}\\
\sum\limits_{y=N}^{\infty}\vol{\chainlet{y}{j}} & \quad \text{if $i = N$ and $j<N$,}\\
\sum\limits_{y=N}^{\infty}\sum\limits_{z=n}^{\infty}\vol{\chainlet{y}{z}} & \quad \text{if $i = N$ and $j = N$.}\\
\end{array} \right.
\] 
\end{definition}

\begin{definition}{Occurrence Matrix}. We denote the total number of chainlet $\chainlet{x}{y}$ in a graph snapshot as $\occ{\chainlet{x}{y}}$. Chainlet counts obtained from the graph snapshot are stored as an $N\times N$-matrix $\mathcal{O}$ such that for $i\leq N,j\leq N$  \[
\mathcal{O}[i,j]=  \left\{
\begin{array}{l l}
\occ{\chainlet{i}{j}} & \quad \text{if $i<N$ and $j<N$, }\\
\sum\limits_{z=N}^{\infty}\occ{\chainlet{i}{z}} & \quad \text{if $i<N$ and $j =N$,}\\
\sum\limits_{y=N}^{\infty}\occ{\chainlet{y}{j}} & \quad \text{if $i = N$ and $j<N$,}\\
\sum\limits_{y=N}^{\infty}\sum\limits_{z=N}^{\infty}\occ{\chainlet{y}{z}} & \quad \text{if $i = N$ and $j = N$.}\\
\end{array} \right.
\] 
\end{definition}
 
\begin{example}
Consider the following example matrices \begin{equation*}
O = 
\begin{bmatrix}
0 & 2 & 1 \\
1 & 1 & 1 \\
1 & 0 & 3
\end{bmatrix},~
A = 
\begin{bmatrix}
0 & 3.58 & 0.5 \\
1.9 & 1.75 & 3 \\
2.8 & 0 & 4
\end{bmatrix}.
\end{equation*}
 If we define $N=2$ for this example, $\mathcal{O}[1,3]$,$\mathcal{O}[2,3]$, $\mathcal{O}[3,1]$, $\mathcal{O}[3,3]$ and $\mathcal{A}[1,3]$,$\mathcal{A}[2,3]$, $\mathcal{A}[3,1]$, $\mathcal{A}[3,3]$ values must be stored inside the $2\times2$ matrices. The updated matrices are 
 \begin{equation*}
O = 
\begin{bmatrix}
0 & 2 +1 \\
1+1 & 1 + 1 +3 \\
\end{bmatrix},~
A = 
\begin{bmatrix}
0 & 3.58+0.5 \\
1.9+2.8 & 1.75+3+4 \\
\end{bmatrix}.
\end{equation*}
\end{example}

\textbf{Extreme Chainlets:} In occurrence and amount matrices, choosing an $N$ value, such as $N=20$, means that a chainlet with more than 20 inputs/outputs (i.e., $\chainlet{x}{y}$ s.t., $ x \geq 20$ or $y \geq 20$) is recorded in the $N$-th row or column. That is, we aggregate chainlets with large dimensions that would otherwise fall outside matrix dimensions. We use the term \textit{extreme chainlets} to refer to these aggregated chainlets on the $N$-th row and column.

Extreme chainlets are large transactions that may involve more than 10 thousand addresses and are also large in coin amounts. We see two behavior in extreme chainlets. In the first case, chainlets are of the type $\chainlet{i<20}{j>20}$; less than 20 input addresses (usually one or two) sell coins to more than 20 addresses. An example is the Bitcoin transaction \href{{www.blockchain.com/btc/tx/a79b970c17d97557357ec0661a2b9de44724440e1c635e1b603381c53ece725d}}{{\tiny a79b970c17d97557357ec0661a2b9de44724440e1c635e1b603381c53ece725d}} in 2018. These extreme chainlets are split-chainlets which may indicate selling behavior.

In the second case, chainlets are of the type $\chainlet{i>20}{j<20}$. This chainlet type is useful in finding large coin buys. Usually, an online exchange collects coins of many individual sellers and creates an extreme chainlet that has few (usually one or two) output addresses. 

\begin{figure}
  \begin{center}
    \includegraphics[width=0.54\textwidth]{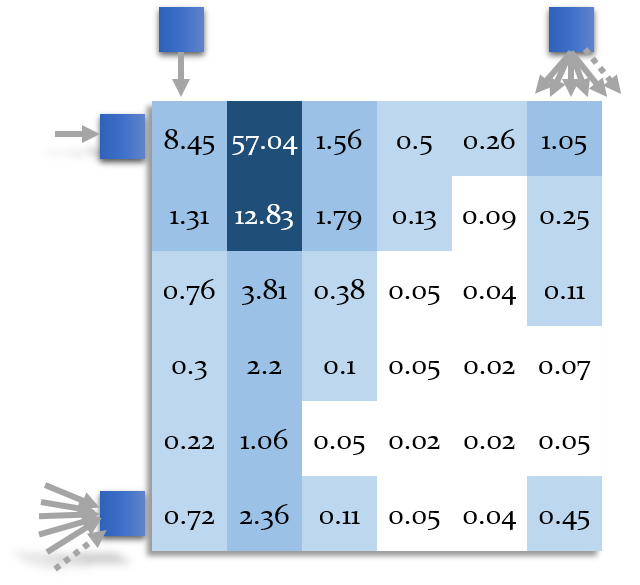}
  \end{center}
  \caption{Percentage of Bitcoin chainlets for $N=6$. Dark colors indicate higher percentages. The first row shows the percentages of one input chainlets. For example, 8.45\% of all Bitcoin transactions are 1-to-1 chainlets, whereas 57.04\% are 1-to-2 chainlets.}
  \label{fig:percentages}
\end{figure}

Extreme chainlets capture large coin movements.  In our research, we have found that information from extreme chainlets shows utility in volatility, risk, and price prediction in cryptocurrencies~\cite{akcora2018EL,deyCSDA2018}.

\section{Ethereum: Account Networks}
Account blockchains, such as Ethereum, do not use the UTXO data structure of Bitcoin. Unlike UTXO coin transactions involving as few as two or as many as thousands of addresses, coin transactions on account blockchains involve only two addresses: sender and receiver.

Many design choices behind account blockchains originate from the most popular account-based blockchain Ethereum. In this section, we will teach account networks by using Ethereum as our data source. However, the reader will find similar, if not the same, network and graph models in other account blockchains.

A major difference between UTXO and account blockchains is the type and variety of networks. UTXO networks are transaction and lightning networks, whereas, on account blockchains, we can observe the following networks:

\begin{enumerate}
    
    \item \textbf{Coin transaction network.} Similar to the UTXO transaction network, this network is created from the coin (ether) transfers between addresses. Network edges only carry the native currency (coin) of the blockchain.
   
    \item \textbf{Token transaction networks.} Asset trading networks that are created by internal smart contract transactions. 
    
    \item \textbf{Trace network.} Interactions between all address types. The name trace implies that a transaction triggers a cascade of calls to smart contracts or externally owned addresses.
 \end{enumerate}

In this section, we discuss the three networks and their building blocks separately. We begin by first outlining network node and edge types.

It is useful to re-consider the Ethereum address types. In account networks, we classify addresses into two node types and one special address. 
\begin{enumerate}
    \item Externally owned address (EOA) has a private key.  An EOA is managed by a real-life entity such as an investor or a centralized exchange. An entity may create and use multiple EOAs at the same time. Typically, an EOA is used over a long time in many transactions.
    \item Smart contract address does not have a private key. We can only distinguish Smart contract addresses by searching for smart contract code at the address. 
    \item Null address \shade{0x000..} node that has two use cases. First, a smart contract is created by using the NULL address as the receiver. Second, the NULL address is used to dispose of crypto assets. Any coin or asset sent to the address cannot be reclaimed and considered burned. The address may have a private key, but finding it by trial is considered impossible.  
\end{enumerate}

\begin{figure}[t]
    \centering
    \includegraphics[width=0.5\linewidth]{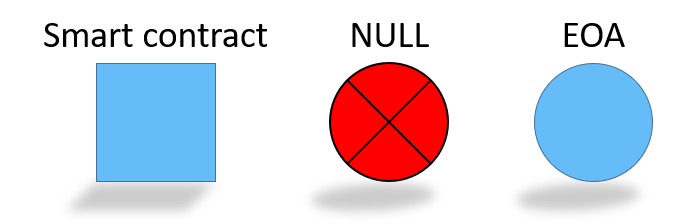}
    \caption{Three address types and corresponding node shapes in account networks. }
    \label{fig:accountaddtypes}
\end{figure}

Figure~\ref{fig:accountaddtypes} visualizes the three-node symbols that we will use to teach account networks. 

\subsection{Account Transaction Network}

The account transaction network contains coin transfers between the three node types. The network edges always start from an EOA because a contract address or a NULL address cannot initiate a transaction.  However, a smart contract can send coins to an EOA when a trace initiates the transfer (we will cover traces shortly). Specifically, transaction network edges are from i) EOA to EOA, ii) EOA to contract, iii) EOA to NULL address. iv) contract to EOA. 

Network edges may have i) coin amount, ii) account nonce, iii) gas price, and iv) timestamp features. 

The coin amount in an edge is in subunits (Wei). Account nonce is a number that orders transactions initiated by an EOA. Miners must mine transactions of an address in nonce order. For example, in Table~\ref{tab:accounttxedges} $a_1$ creates four transactions with nonce values 0 to 3. A future transaction of $a_1$ with nonce 5 has to wait because the transaction of nonce 4 has not been mined. Nonce order ensures that the network cannot have out-of-order or missing edges. However, miners can mine multiple transactions from an EOA in the same block according to the nonce order. 

\begin{figure}[t]
  \begin{center}
    \includegraphics[width=0.48\textwidth]{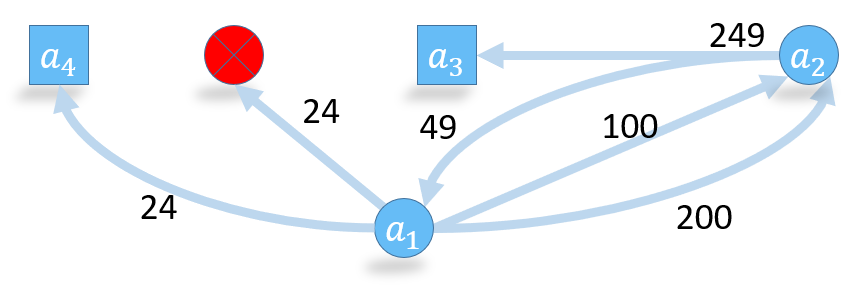}
  \end{center}
  \caption{A toy network of five nodes and edges with ether amounts (in wei) from four blocks. Edge features are given in Table~\ref{tab:accounttxedges}. The address $a_1$ owns 300 weis from an earlier transaction (not shown). Transaction fee is one wei. }
  \label{fig:accounttxnetwork}
\end{figure}

Although we mention transaction time as an edge feature, transactions do not have timestamps themselves. The timestamp comes from the timestamp of the block that contains the transaction. Transactions of a block are ordered in the block by the miner, and online explorers record the order as the \textit{block index}.  However, even the miner cannot truly know when a transaction was created.  

Another edge feature could be the \textit{used gas} field of the transaction. However, the feature is not useful for the transaction network because the gas cost equals the base fee (currently 21000 gas) for coin transfers. The input data field of the transaction is empty for coin transfer transactions. 

Figure~\ref{fig:accounttxnetwork} shows an example account transaction network. Table~\ref{tab:accounttxedges} gives the edge table of this network.

We easily model the account transaction network as a directed and weighted multigraph (i.e., multiple edges exist between node pairs). Although address creation is cheap, address reuse is not discouraged on Ethereum, and a node may appear in multiple blocks and send and receive coins for an extended period. Furthermore, smart contracts have permanent addresses. These factors help us track node behavior and network dynamics in time.

\begin{table}[b]

    \centering
    \small{
    \begin{tabular}{c c c c c c c}
     \toprule
   block height& from & to & amount (wei) &  nonce & block index&timestamp\\
   \midrule
     10646423&    $a_1$ & $a_2$ & 100 & 0 &1 &Aug-12-2020 05:11:17 PM +UTC\\
     10646423&    $a_1$ & $a_2$ & 200 & 1 &2&Aug-12-2020 05:11:17 PM +UTC\\
     10646424&    $a_2$ & $a_3$ & 249 & 0 &1&Aug-12-2020 05:11:18 PM +UTC\\
     10646424&    $a_2$ & $a_1$ & 49 & 1 &2&Aug-12-2020 05:11:18 PM +UTC\\
     10646425&    $a_1$ & NULL  & 24 & 2 &1 &Aug-12-2020 05:11:26 PM +UTC\\
     10646426&    $a_1$ & $a_4$ & 24 & 3 &1&Aug-12-2020 05:11:43 PM +UTC \\
     \bottomrule
    \end{tabular}
    }
    \caption{Network edges and their features of some transaction from blocks 10646423 to 10646426.}
    \label{tab:accounttxedges}
\end{table}

\subsection{Token Transaction Networks}

A token is created by deploying a smart contract where the token's features and business logic are defined. Any blockchain participant can create a token and facilitate its trade. The token's contract defines meta attributes about the token, such as its symbol, total token supply, or decimals. Only the address of a token is unique in the blockchain; multiple tokens can have the same symbol, which creates confusion in trades.

Currently, there are more than a hundred thousand tokens on Ethereum~\cite{lee2019measurements}. We consider each token to have its network and set of traders. 

A token's current supply is the number of token instances created by the smart contract. If the instances are given unique ids, the token is considered non-fungible (as in the ERC 721 standard); each instance has its characteristics, owner, and price. Otherwise, one token instance will be equal to any other; the token is considered to be fungible (as in the ERC 20 standard). In this section, we will consider networks of fungible tokens, but we can easily extend our analysis to the non-fungible case.

A token transaction network has EOA, NULL, and smart contract addresses as nodes. We outline the following three types of transactions that a Data Scientist must know to analyze token networks.

\begin{itemize}
    \item The creation transaction that assigns an address for the token initializes its smart contract and state variables.
    \item A trade transaction that moves some tokens between addresses.
    \item A management transaction that can only be initiated by the smart contract creator (or any address that the owner specifies). The transaction may delete the contract or forward its balance (in ether or token) to another address. We will study management transactions in Section~\ref{sec:tracenetwork}.
\end{itemize}

\begin{table}[b]
    \centering
    \begin{tabular}{l c c c c c c}
     \toprule
  token &block height&from & to & amount (Gwei) & tx fee (Gwei)& input data\\
   \midrule
    Binance: BNB Token&3978343 &    \tiny{0x00c5...454} & \tiny{0xb8c7...d52} & 0 & 32643080 &[.....]\\
    ChainLink: LINK Token&4281611 &    \tiny{0xf550...780} & \tiny{0x5149...6ca} & 0 & 21895728 &[.....]\\
    Tether USD &4634748 &    \tiny{0x3692...d57} & \tiny{0xdac1...ec7} & 0 & 12683176 &[.....]\\
    
     \bottomrule
    \end{tabular}
    \caption{Contract creation transactions for three popular tokens. The input data field contains the contract code. The from address is the owner of the smart contract, which resides at the \textit{to} address. The Ethereum protocol deterministically computes the address from i) the owner's address and ii) the account nonce when the transaction is created (by the owner).}
    \label{tab:accountcontractcreation}
\end{table}

Table~\ref{tab:accountcontractcreation} lists the creation transactions of three popular tokens on Ethereum. The from field lists the contract creator; this address is the owner of the token. Contract creation is expensive, and the transaction fee increases with more elaborate contracts. The contract code appears in the input data field of the transaction. 

\begin{remark}
 Usually, the transaction that creates a smart contract carries no ether (as shown with 0 amounts in Table~\ref{tab:accountcontractcreation}). The actual payload is the smart contract code written in the input data field of the transaction. 
\end{remark}
 
Token networks evolve with changing user balances that we denote as edges between addresses. However, token transfers are \textit{internal transactions} which are not broadcast to the network in the form of an ordinary Ethereum transaction. In other words, what we call a token trade is, in reality, an update of balances in smart contract variables of the token. It is as simple as changing the values of two keys in a hashmap.

\begin{figure}[t]
\begin{subfigure}{.3\textwidth}
  \centering
  \includegraphics[width=.8\linewidth]{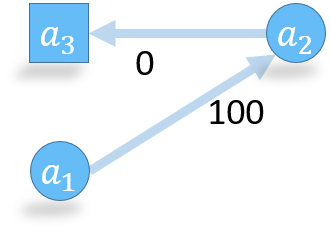}
  \caption{Two transactions are mined in blocks.}
  \label{fig:tokentrade1}
\end{subfigure}
~
\begin{subfigure}{.3\textwidth}
  \centering
  \includegraphics[width=.8\linewidth]{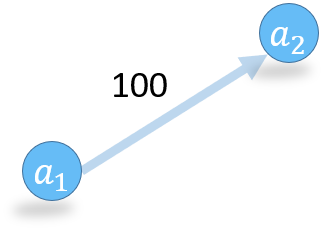}
  \caption{Account transaction network records an edge for transferring 100 wei}
  \label{fig:tokentrade2}
\end{subfigure}
~
\begin{subfigure}{.3\textwidth}
  \centering
  \includegraphics[width=.8\linewidth]{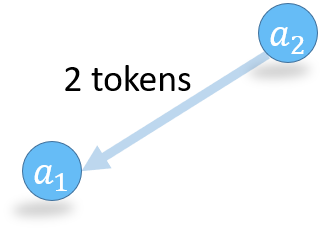}
  \caption{Token transaction network records an edge for transferring two tokens}
  \label{fig:tokentrade3}
\end{subfigure}
\caption{A trade between addresses $a_1$ and $a_2$ for the tokens issued by $a_3$. Transaction and token networks capture partial views from the transactions given in (\subref{fig:tokentrade1}). The zero amount transaction is omitted in the token transaction network, whereas the transfer of two tokens are not captured in an Ethereum transaction network. }
\label{fig:tokentrade}
\end{figure}

We will explain the observed transactions and states with the help of Figure~\ref{fig:tokentrade}. A token trade involves the following steps.
\begin{enumerate}
    \item Address $a_2$ owns tokens that are issued by $a_3$. $a_2$ could have received tokens through various channels. For example, $a_2$ may have bought them in an earlier transaction, or the tokens might have been given to $a_2$ by the owner of $a_3$.
    \item A price per token is agreed upon between $a_1$ or $a_2$ by any off-the-blockchain channel or through a clause in the contract (e.g., the contract sets token price as 50 Wei).
    \item $a_1$ pays $a_2$ an ether amount (100 Wei in Figure~\ref{fig:tokentrade1}) by creating a transaction, where $a_2$ is the receiver and the input data field of the transaction is blank. Here $a_1$ sends the transaction to the network to be mined.
    \item $a_2$ downloads the latest blocks and sees that 100 Weis have been sent to its address from $a_1$. In turn, $a_2$ uses the conversion rate and decides to send 2 tokens to $a_1$.
    \item \label{step:toknetrade} $a_2$ creates a transaction where the receiver is $a_3$, and the transferred ether amount is 0. However, the input data lists $a_1$ as the recipient of two tokens. Typically the input data is a call to the \textit{transfer} function of the smart contract at $a_3$ with two parameters: to\_address:$a_1$ and amount:$2$. $a_2$ sends this transaction to the network to be mined.
    \item At every node of the Ethereum network, the Ethereum Virtual Machine executes the transaction of step~\ref{step:toknetrade}. The smart contract at $a_3$ decreases the balance of $a_2$ by two tokens and increases the balance of $a_1$ by two tokens. This balance update, which is an internal transaction, is recorded as an edge from $a_2$ to $a_3$ in the token network of Figure~\ref{fig:tokentrade3}, but not sent to the network as a transaction. This step fails if $a_2$ does not own two tokens. 
    \item $a_1$ downloads and observes the transaction from $a_2$ to $a_3$. The node at $a_1$ can run the transaction in its local Ethereum Virtual Machine to ensure that tokens are transferred without any error. 
    
\end{enumerate}

As Figure~\ref{fig:tokentrade} shows, transaction and token networks have different views from the two mined transactions. Without running a virtual machine and executing the transactions, we cannot observe the internal transactions nor create the token transaction network.

Token networks attract trades and traders every day. As Ethereum addresses (both EOA and contract) are reused in multiple days, a token network may see the same node trading in multiple days. However, daily token networks are sparse and consist of disconnected components, and very few traders appear in a token's network every day. 

Some traders appear in networks of multiple tokens, making token networks an ideal setting for studying multi-layer networks. Each token network constitutes a layer with edges and nodes, and nodes overlap between layers.

\subsection{Trace Network}
\label{sec:tracenetwork}
Ethereum stores an ecosystem of addresses, smart contracts, and decentralized organizations. In transaction and token networks, we studied financial relationships between addresses. This section now shifts our focus to relationships, call-dependencies, inheritances, and other interactions between Ethereum addresses. A trace network stores these relationships where nodes are EOA and smart contract addresses, and edges are interactions between EOA-contract, contract-EOA, and contract-contract pairs.

An EOA creates a transaction that is directed to the address of a smart contract. The smart contract can execute a function and terminate or call functions of other smart contracts. The contract can also move coins to EOAs. The trace can be extended until the transaction gas limit is exhausted. All the addresses that are involved in this call/interaction chain create a \textit{trace}. In graph terms, a trace is a hyper-edge (i.e., an edge that connects more than two nodes). 

A trace can involve every operation that a smart contract can execute. For example, a trace can create a smart contract, call a smart contract function, or delete a smart contract. As such, we can label parts of the trace with the executed operation.

\begin{figure}[ht]
    \centering
    \includegraphics[width=0.9\linewidth]{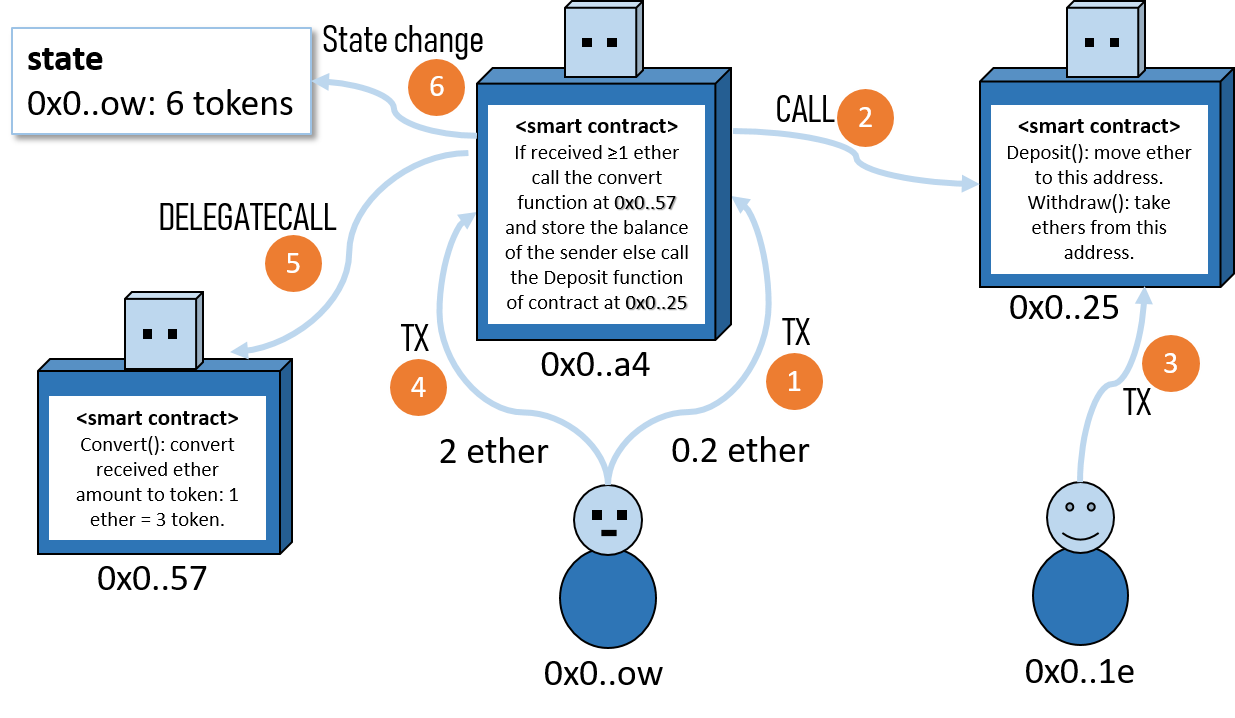}
    \caption{Transactions and message calls. The account \shade{0x0..ow} initiates the first transaction (1st step) whose amount is forfeited as it is less than 1 ether. The forfeited amount is deposited to the contract at \shade{0x0..25} for the owner (\shade{0x0..1e}) of the smart contract to withdraw it in the third step. Next, the account creates another transaction (4th step) that successfully converts 2 ethers to 6 tokens. }
    \label{fig:tracecalls}
\end{figure}

\begin{example}
We construct a trace network from a toy scenario given in Figure~\ref{fig:tracecalls}. We show addresses with their first and last characters (e.g., \shade{0x0..57}). 
Figure~\ref{fig:tracecalls} contains three smart contracts (\shade{0x0..57}, \shade{0x0..a4} and \shade{0x0..25}) and two externally owned accounts (\shade{0x0..ow} and \shade{0x0..1e}). Edges denote transactions or message calls. Transactions are created by externally owned accounts explicitly and mined in blocks, whereas message calls are not.

In this example, a contract acts as a salesman for a cryptoasset:  \shade{0x0..a4} receives ethers from addresses, and stores their amount of token in its storage. The contract has a rule that the minimum deposit amount is 1 ether. \shade{0x0..ow} is not aware of this rule at first and creates a transaction that sends 0.2 ethers to \shade{0x0..a4}, which calls a deposit function at \shade{0x0..25} that seizes the amount and stores it in the address. \shade{0x0..1e} can observe \tikz[baseline=(char.base)]{            \node[shape=circle,fill=orange,draw,inner sep=2pt] (char) {1};} in the blockchain as a transaction, but \tikz[baseline=(char.base)]{            \node[shape=circle,fill=orange,draw,inner sep=2pt] (char) {2};} is a message call that requires \shade{0x0..1e} to execute the contract call to discover. 

Usually contracts have withdraw functions that allow the contract creator to remove the deposited ethers from the contract.  In \tikz[baseline=(char.base)]{            \node[shape=circle,fill=orange,draw,inner sep=2pt] (char) {3};} \shade{0x0..1e} makes a contract call transaction to withdraw these ethers. 

\shade{0x0..ow} learns about the minimum amount rule, and sees that its 0.2 ethers are forfeited. \shade{0x0..ow} makes yet another attempt to buy tokens by sending 2 ethers again in transaction \tikz[baseline=(char.base)]{            \node[shape=circle,fill=orange,draw,inner sep=2pt] (char) {4};}. This time, \shade{0x0..a4} accepts the amount, and DELEGATECALLS a conversion function from \shade{0x0..57}. which directly stores a state variable that records account \shade{0x0..57} has 6 tokens (three for each ether). 

So far, three transactions (\tikz[baseline=(char.base)]{            \node[shape=circle,fill=orange,draw,inner sep=2pt] (char) {1};}, \tikz[baseline=(char.base)]{            \node[shape=circle,fill=orange,draw,inner sep=2pt] (char) {4};} and  \tikz[baseline=(char.base)]{            \node[shape=circle,fill=orange,draw,inner sep=2pt] (char) {3};}) have been initiated by EOAs and mined in blocks. We observe the following three traces from the three transactions:

\begin{enumerate}
    \item Trace 1: \shade{0x0..ow} $\xrightarrow{ \tikz[baseline=(char.base)]{            \node[shape=circle,fill=orange,draw,inner sep=2pt] (char) {1};}}$ \shade{0x0..a4}$\xrightarrow{ \tikz[baseline=(char.base)]{            \node[shape=circle,fill=orange,draw,inner sep=2pt] (char) {2};}}$ \shade{0x0..25}
    \item Trace 2: \shade{0x0..1e} $\xrightarrow{ \tikz[baseline=(char.base)]{            \node[shape=circle,fill=orange,draw,inner sep=2pt] (char) {3};}}$ \shade{0x0..25}  
    \item Trace 3: \shade{0x0..ow} $\xrightarrow{ \tikz[baseline=(char.base)]{            \node[shape=circle,fill=orange,draw,inner sep=2pt] (char) {4};}}$ \shade{0x0..25}$\xrightarrow{ \tikz[baseline=(char.base)]{            \node[shape=circle,fill=orange,draw,inner sep=2pt] (char) {5};}}$ \shade{0x0..57}
\end{enumerate}

Note that \tikz[baseline=(char.base)]{            \node[shape=circle,fill=orange,draw,inner sep=2pt] (char) {6};} creates a state change (an internal transaction) but it is not recorded in the trace network.

\begin{figure}[t]
    \centering
    \includegraphics[width=0.7\linewidth]{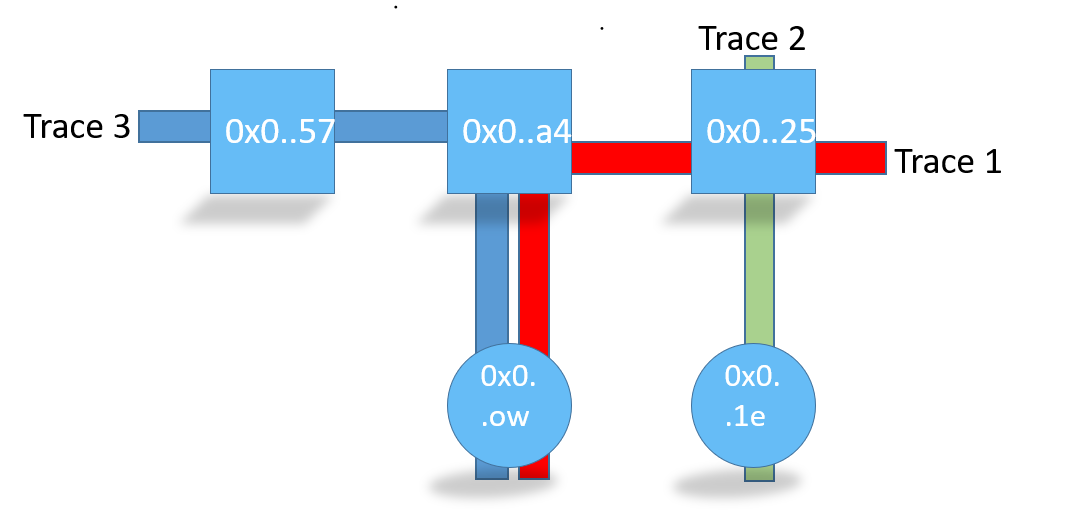}
    \caption{A hyper-graph from Ethereum traces. Three hyper-edges connect five nodes. Nodes that appear on a trace are all considered neighbors. A finer grained hyper graph can add edge features based on interaction types between node pairs. }
    \label{fig:tracenetwork2}
\end{figure}

Placing all these traces (hyper-edges) together, we obtain the hypergraph shown in Figure~\ref{fig:tracenetwork2}. 

\end{example}

\begin{remark}
 The Ethereum virtual machine executes every transaction in order and executes calls accordingly. There is no concurrency in Ethereum transaction execution; EVM does not make two calls simultaneously. When a contract calls two contracts sequentially, EVM will make the second call only after the first contract call (and any further contract calls it makes) ends. 
\end{remark}

In theory, traces can reach large sizes in the shape of trees with many branches. However, the transaction fee grows with additional calls and prevents the creation of huge traces. Note that even when a transaction sets a high gas limit to pay a big transaction fee, the transaction will not be mined if the gas used exceeds the Ethereum block gas limit (currently 12.5K). If a trace creates a call loop, this may deplete smart contract balances before the transaction gas limit is reached.

We can understand smart contract behavior by trace network analysis. For example, we can study the network to discover unexpected call branches in smart contracts.

\section{Ripple: Credit Networks}
Ripple (\url{www.ripple.com}) and Stellar (\url{www.stellar.org}) are two credit networks that closely resemble the ancient Hawala system (in Arabic, hawala means to transfer or trust). The main idea of Hawala is to allow a money sender to use connections of people who trust each other to make a payment in a distant geographical location. Merchants have historically used these types of networks to transfer money between countries~\cite{el2003informal}. 
 
We model credit networks as directed, weighted graphs that we build from \textit{trust lines} between address pairs. A trust line is a directed edge between two addresses, which implies that the source address trusts the target address. We can add weight (i.e., money amount) to the edge to show the limits of that trust. In a credit network built from trust lines, a directed edge is a promise by the source node that it will let the target node use a loan amount in a future transaction. Trust lines can be deleted or updated for amounts in time. 

It is helpful to explain a few confusion points in Ripple to a blockchain researcher. Ripple uses the term \textit{ledger} instead of a block. Transactions have multiple types and may involve financial constructs (e.g., checks), user-issued currencies (e.g., USD), and path-based settlements. The reader should remember that Ripple contains elaborate business logic in its building blocks (e.g., transactions). Before we delve into Ripple, we will use a few scenarios to explain how we can use a network of trust lines to make a payment. 

\begin{remark} Academic articles have conflicting views on how to represent a trust line as a directed graph edge.  Here, we follow the notation used in the official Ripple documentation, i.e., the edge is from the lender (source) to the borrower (target). The lender trusts the borrower.
\end{remark}
\begin{figure}[!h]
    \centering
    \includegraphics[width=\linewidth]{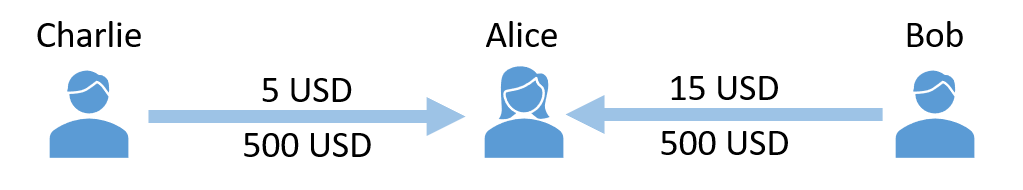}
    \caption{Two Ripple trust lines between Charlie, Bob, and Alice. Charlie has created a trust line for Alice for 500USD. Alice has already used 5USD from the trust line. The amount below both edges, 500 USD, is the maximum limit (set by Charlie and Bob) that Alice can use.}
    \label{fig:rippleedge}
\end{figure}

Figure~\ref{fig:rippleedge} shows two trust lines between Alice, Charlie, and Bob. Each edge has two edge features. First, a value below the edge (500 USD for both edges) is the maximum credit amount (known as the \textit{limit}) that Charlie and Bob trust with Alice that Alice can use in future transactions. Second, a value above the edge (5 and 15 USD) is the \textit{balance} amount that Alice has already used with Charlie and Bob. For example, Alice can use $500-15=$485 in a future transaction by using the trust line from Bob.

\begin{figure}
\begin{subfigure}{.5\textwidth}
  \centering
  \includegraphics[width=.95\linewidth]{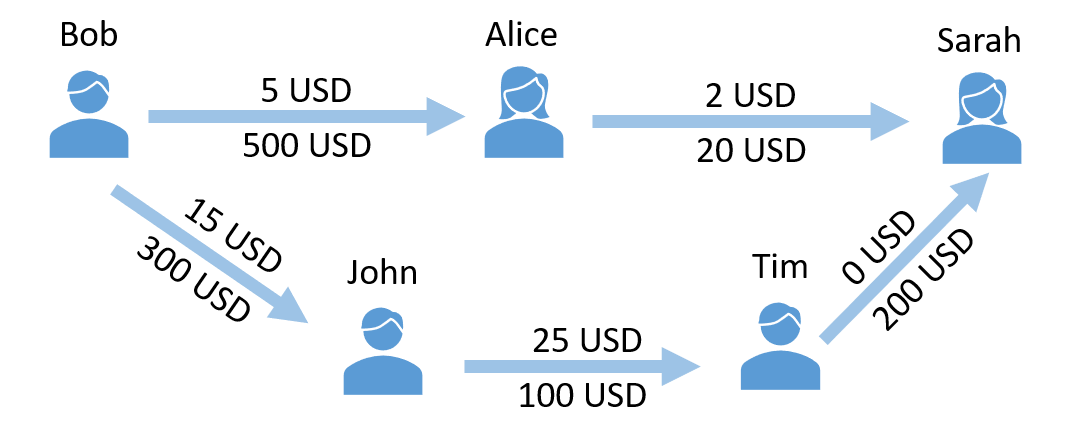}
  \caption{Trust lines. Before Sarah sends any currency, the Ripple trust graph includes five nodes and five trust lines. }
  \label{fig:rippletrustnetwork1}
\end{subfigure}%
~
\begin{subfigure}{.5\textwidth}
  \centering
  \includegraphics[width=.95\linewidth]{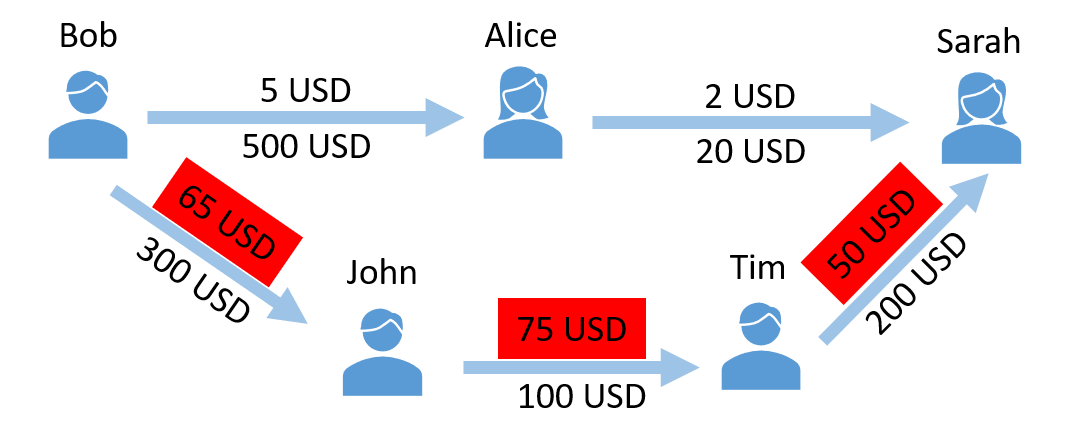}
  \caption{Payment. After the rippling, owed amounts are updated and shown in red rectangles.}
  \label{fig:rippletrustnetwork2}
\end{subfigure}
\caption{An example of the Ripple trust graph before (left) and after (right) Sarah uses rippling to send 50 USD to Bob. On the (left) network, Sarah has two paths that connect her to Bob: 1) Bob $\rightarrow$ Alice $\rightarrow$ Sarah and 2) Bob $\rightarrow$ John $\rightarrow$ Tim $\rightarrow$ Sarah. However, in the 1st path that traverses two trust lines, the trust line between Sarah and Alice has a max limit of 20 USD (2 USD of which is already used), which is less than the 50 USD that Sarah wants to send. As a result, Sarah cannot use the first path to send all 50 USD. The second path is longer than the first because it traverses three trust lines; however, the limits allow the 50 USD transfer. Sarah uses the second path for rippling.  Owed amounts (values above edges) are increased by the transferred amount: 50 USD. Sarah needs to consider updated (right) values in future transactions; she can no longer send 50 USD to Bob because the trust line John $\rightarrow$ Tim has a current capacity of USD 100 - USD 75 = USD 25 only. The updated (increased) amounts imply that after the payment, Sarah owes 50 USD to Tim, who owes 75 USD to John, who owes 65 USD to Bob.}
\label{fig:rippletrustnetwork}
\end{figure}

Users may create the Ripple trust line due to two reasons. First, Alice may have paid 500 USD to Charlie and Bob each in an off-the-Ripple transaction. Second, Charlie and Bob may have trusted Alice based on her reputation and created trust lines for her to use. In practice, the amounts are almost always due to the first reason, i.e., offline deposits. By default, Ripple limits are lower-bounded by 0 and upper-bounded by $\infty$. A high limit creates credit risks for the lender. We make a Ripple trust graph from such trust lines.  We can delete an existing trust line by setting its trust limit to zero.
We can use trust lines to make payments to third addresses over a path that traverses individual trust lines in the same currency, such as USD. Traveling trust lines is known as \textit{rippling}. Figure~\ref{fig:rippletrustnetwork} will help the reader learn basic rippling across trust lines.

\subsection{Ripple Networks}

On cryptocurrencies, a single type of coin, such as bitcoin, is traded between addresses. Similarly, Ripple has its currency called XRP with a \textit{drop} sub-unit (1 XRP = 1 million drops). However, Ripple also allows users to issue currencies and create trust lines in the issued currencies.

Ripple officially recognizes two main currency types: i) Ripple (abbreviated as XRP) and ii) user-issued currencies. Academic articles further divide user-issued currencies into country currencies (e.g., US dollar, European Euro, Japanese Yen), cryptocurrencies (e.g., Bitcoin), and fictional currencies such as tokens. For example, in Figure~\ref{fig:rippleedge}, USD amount is a user-issued currency.  User-issued digital tokens are fictional currencies with no outside backing. Traders must be aware that such tokens have no inherent value.

XRP is issued natively and has no issuer. All other currencies are represented as \textquote{currency.issuer} in transactions. We represent a currency with i) three characters such as USD, EUR, or ii) less frequently by a 40 character string. USD issued by address1 is USD.address1 on the Ripple ledger (also called the XRP Ledger). Multiple issuers can issue a currency: USD.address1 and USD.address2 are considered two different currencies, even though they are both USD currencies. However, in rippling user issued currencies of the same denomination (e.g., USD) are considered to be the same currency. Trust in the issuer determines the fungibility of a user-issued currency. If Alice does not trust address1 but trusts address2, 1 USD.address1 will not be worth 1 USD.address2 for Alice. Accordingly, Alice may refuse to participate in rippling transactions that bring her USD.addres2 amounts. 

Each Ripple address needs to store a non-trivial XRP amount as a reserve that it cannot spend; otherwise, Ripple considers the address deleted. The reserve requirement (currently 20 XRP) discourages multiple address creation and usage. If a security breach occurs, a user may change its address; however, best practice reuses the same address for multiple transactions.

\begin{info}
A currency issuer may set a transfer fee for every transaction in that currency between other addresses. The fee is similar to earning a commission for every transaction in the currency. The issuer of a popular currency thus accumulates transfer fees, which reduces the amount of offline liability it must hold to exist as an entity in the Ripple network. \end{info}

The issuer can also freeze all transactions of its currency (XRP cannot be frozen by anyone). However, holders of the currency can still redeem the currency by returning it to the issuer and getting paid outside of the XRP Ledger. However, in real life, the issuer may have already gone bankrupt or may ignore redeeming requests.  Every address on the ledger must make its own trust decisions by considering these issues and create/allow trust lines accordingly. 

\begin{figure}
    \centering
    \includegraphics[width=0.9\linewidth]{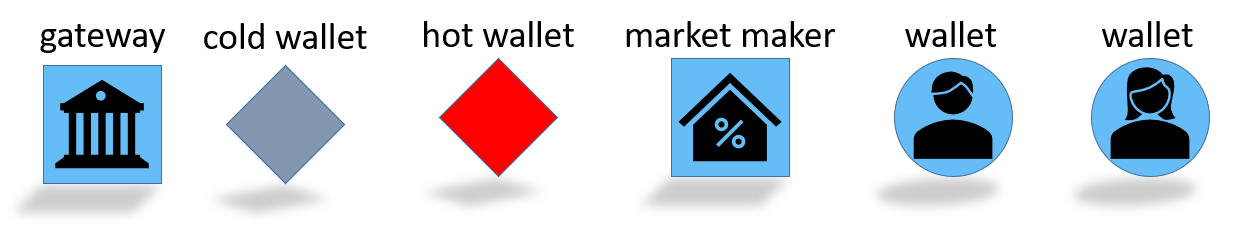}
    \caption{Ripple address types.}
    \label{fig:rippleaddress}
\end{figure}

We outline Ripple Ledger address types in Figure~\ref{fig:rippleaddress} as a gateway, cold wallet (issuing address), hot wallet (operational addresses), market maker, and wallets (i.e., ordinary traders). Gateways are the currency issuers of the Ripple network and link the XRP Ledger to the rest of the world. All network nodes can buy Ripple or any other issued currency by paying money (e.g., US dollars) to the gateway (outside the Ripple ledger).  The following steps, taken verbatim from \href{https://xrpl.org/issued-currencies-overview.html}{the Ripple documentation}, explain the six necessary steps.  

\begin{enumerate}
    \item A customer sends money to a gateway's offline accounts. This could be fiat money, Bitcoin, or any other asset not native to the XRP Ledger.
\item The gateway takes custody of the money and records it.
\item The gateway issues a balance in the XRP Ledger, denominated in the same currency, to an address belonging to the customer. This is done by creating a trust line to the customer's Ripple address.
\item The customer uses the issued currency in the XRP Ledger, such as by sending cross-currency payments or trading in the decentralized exchange.
\item A customer (not necessarily the one who deposited the money initially) sends the issued currency to the gateway's XRP Ledger address.
\item The gateway confirms the customer's identity who sent the balance in the XRP Ledger funds and gives the corresponding amount of money outside the XRP Ledger to that customer.
\end{enumerate}

An institution may use both cold and hot wallets on the XRP ledger. Connected to the Internet, a hot wallet signs ordinary transactions, and the cold wallet stores currencies offline and is used less frequently.

 \begin{figure}
  \begin{center}
    \includegraphics[width=0.53\textwidth]{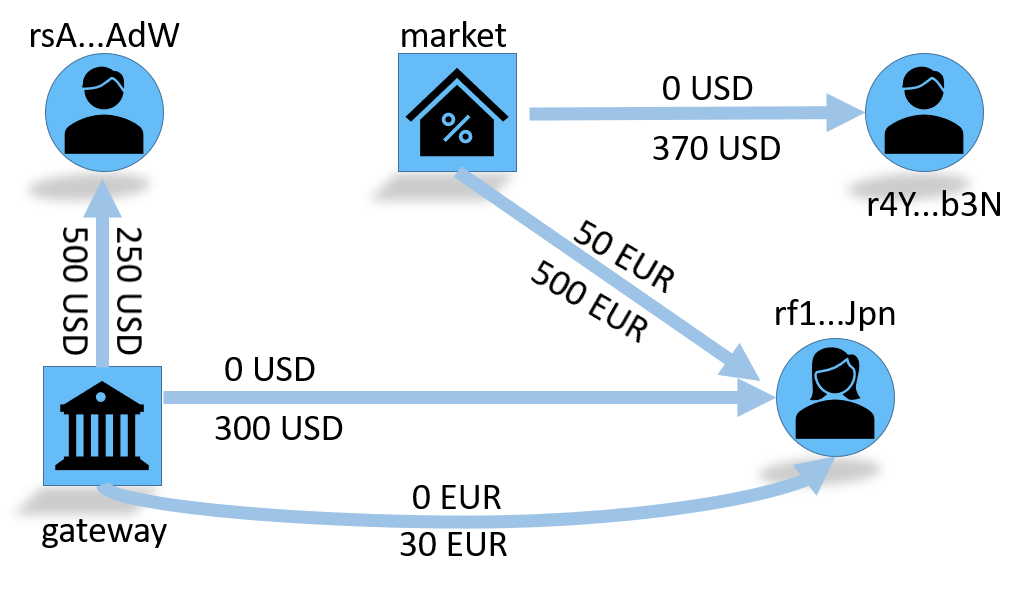}
  \end{center}
  \caption{A Ripple trust graph. The market has trust lines to two nodes. Note that it is the gateway and market that have credit lines to the users. If the balance (edge weight above the line) is non-zero, the node has already used the credit to make a payment to another node. Assume that market and gateway are low accounts. Below lines, we show low and high limits, respectively.}
   \label{fig:ripplemultipleedge}
\end{figure}

A market maker is an address that converts a currency to another and takes a conversion fee through its \textit{offer}, which is broadcast to the ledger in an \textit{offer create} transaction. The offer may immediately consume earlier offers and gain the market maker what currency it desires. If the offer is not fully consumed, it is stored in the ledger as an \textit{offer object} that can be used in future payments or consumed by future offers (of other addresses). Offers must be updated in real-time as currency conversion rates change (for country currencies, the conversion rate must be updated according to international exchange rates). A standing exchange offer is canceled in an \textit{offer cancel} transaction.

We will teach Ripple networks in two subsections: Ripple trust graph and Ripple payment graph. The trust graph stores trust lines and balances. The payment graph utilizes the trust graph to process transactions.

\subsection{Ripple trust graph}

%\epigraph{A trust line represents an explicit statement of willingness to hold gateway debt obligations. In other words: \textquote {I'll allow you to owe me up to this much money outside the XRP Ledger.}} {Ripple Documentation (\url{https://xrpl.org/issued-currencies-overview.html}) }

We create the Ripple trust graph from trust lines issued by addresses. In our basic examples of Figures~\ref{fig:rippleedge} and \ref{fig:rippletrustnetwork}, trust lines were not reciprocal. However, in Ripple, both addresses can create a trust line to each other in a currency. Furthermore, two addresses can have trust lines in multiple currencies. 

Ripple creates a single \textit{RippleState} object to connect two accounts per currency, sorts the account addresses numerically, and labels the numerically lower address as \textit{low account} and the other as the \textit{high account}. Assume that Alice and Bob have trust lines in USD and EUR amounts. Ripple creates two RippleState objects for Alice and Bob; one for EUR and one for USD.

The object has high and low limit fields to record the limit from high $\rightarrow$ low and low $\rightarrow$ high trust lines. However, the net balance is a single value shared by the addresses, and Ripple stores the net balance of the trust line from the low account's perspective. 

\begin{remark}
So far, we have used trust lines between Ripple users (e.g., John $\rightarrow$ Sarah) in our payment scenarios. However, ordinary users do not issue currencies on Ripple, which means that actual trust lines connect currency issuer-user pairs (gateway $\rightarrow$ Sarah). 
\end{remark}

We can check the currency issuer in a RippleState with balance. If the balance is positive, the high account is the issuer. If the balance is negative, the low account is the issuer. Often, the issuer has its limit set to 0, and the other account has a positive limit, but this is not reliable because limits can be increased or decreased without affecting an existing balance.

The Ripple trust graph is a directed, weighted multi-graph where an edge has four features: currency name (i.e., type of the edge), balance, low limit, and high limit. Multiple currency edges can connect node pairs. For example, in Figure~\ref{fig:ripplemultipleedge} the gateway and \shade{rf1...Jpn} have trust lines in USD and EUR currencies. We give the graph data of this figure in Table~\ref{tab:rippledata} where we have not excluded 0 high limits.

 \begin{table}
    \centering
    \begin{tabular}{c c cc  c c}
    \toprule
    low&high&currency&balance&low limit&high limit\\
    \midrule
    gateway&rSA...Adw & USD & 250 & 500 & 0  \\
    gateway& rf1...Jpn& USD&0 &300  &0 \\
    gateway& rf1...Jpn&EUR&0 &30& 0  \\
   market& rf1...Jpn&EUR&50 & 500& 0  \\
   market& r4Y...b3n&USD&0 & 370& 0  \\
    \bottomrule
    \end{tabular}
    \caption{Ripple trust graph data. Gateway and market are assumed to be low accounts, whereas user addresses are high accounts. High limit values are 0 because high accounts are all user addresses.}
    \label{tab:rippledata}
\end{table}

A trust line deactivates in three cases. First, the lender may remove the trust line by updating trust parameters in a TrustSet transaction. Second, the lender may freeze the trust line temporarily. Third, the currency issuer can freeze all transactions of the currency (except for the redeeming transaction). In two cases, a node may hold a balance on a trust line greater than the limit, that is, when the node acquires more of that currency through trading and when the node decreases the limit on the trust line. The average node degree is less than four in the trust graph, and the clustering coefficient is around 0.10.

\subsection{Ripple payment graph}

The Ripple trust graph shares edge features with the Ripple payment graph. Additionally, the payment graph stores a node feature called reserve (i.e., XRP balance of the node). We cannot use the reserve amount in transactions. In addition to the base reserve, the reserve amount increases for every object (such as a trust line) that we create. The reserve increase applies only to the node extending trust, not to the node receiving it.

We create the Ripple payment graph from 1) direct payments, 2) path-based settlements, and 3) a few other financial constructs. We will teach these payment types before giving a graph model for the payment graph.

\subsubsection{Direct payments}

A direct payment is an XRP transfer between two Ripple addresses and does not require a trust line. Regardless of what trust lines exist in the trust graph, two ripple addresses can send to and receive XRP from each other. 

The XRP balance of an address must not fall below the total required reserve. Otherwise, the address is not allowed to create some types of transactions. When an address's XRP balance falls below the reserve, the address may issue OfferCreate transactions to buy more XRP or other currencies on its existing trust lines. However, these transactions cannot create credit that we can use to buy XRP to satisfy the reserve requirement  (i.e., when the account balance$<$ the required reserve). For example, the address may pay a transaction fee and create an offer for XRP on the ledger. Once a buyer consumes the offer, the address will receive its XRP payment and satisfy the reserve requirement. 

 When an address is below the reserve requirement, it can send new OfferCreate transactions to acquire more XRP or other currencies on its existing trust lines. These transactions cannot create new trust lines, nor Offer objects in the ledger, so they can only execute trades that consume existing Offers.

\subsubsection{Path-based settlements} 

A path-based settlement transaction connects a source and a destination node pair via a network path through rippling.  For example, in Figure~\ref{fig:ripplemultipleedge} the node \shade{rsA...AdW} can make a payment to \shade{rf1...Jpn} through the gateway. We can make a payment by consuming exchange offers along the path. For example, if the sender wants to make a payment in USD, the path can use trust lines that follow USD$\rightarrow$EUR$\rightarrow$USD, where the arrow indicates a currency exchange. Traversing exchanges is called bridging. If the transaction offers a better cost, currency$\rightarrow$XRP and XRP$\rightarrow$currency conversions can automatically be executed on the path as well. This is called \textit{auto-bridging}.

A path-based settlement can explicitly write one or more paths (called a pathset) in a payment transaction. All paths in a pathset must start and end with the same currency, and the sender can also leave path selection to Ripple. 

Pathfinding is difficult because user XRP balances and trust lines change every few seconds as new ledgers (i.e., blocks of transactions) are validated. As a result, Ripple is not designed to compute the absolute best path. The simplest possible path to connect the steps of the transaction is called the \textit{default path}. In addition to the paths written in the transaction, Ripple can choose to use the default path. The sender must set the tfNoDirectRipple flag to avoid the default path and force Ripple to use the paths of the pathset. 

In a path-based settlement, we can set a few fields of the transaction to restrict path selection. These are the Account (sender), Destination (receiver), Amount (currency and amount to be delivered), and SendMax (currency and amount to be sent, if specified) variables of the transaction. The variables shape the path in these ways: 

\begin{itemize}
    \item The Account (sender) is set to be the first step of the path.
    \item If the SendMax variable is set to be a currency issuer, the issuer must be the second step of the path.
    \item If the Amount variable is set to be a currency issuer, the issuer must be the second-to-last step of the path.
    \item The path ends at the Destination address. 
\end{itemize} 

The actual cost of a path-based payment can change between submission and execution of a transaction based on updated information (e.g., the limit of a trust line) from transactions.

A path-based payment can fail for multiple reasons; first, the receiving address may not have the reserve XRP amount. Ripple checks whether the payment would deliver enough XRP to meet the reserve requirement in such a case. If not, the payment fails. Next, the receiving address may have limitations on receiving payments, such as Deposit Authorization or RequireDest (We will cover these constructs shortly). \footnote{Currency issuers set such rules.} Third, the paths that the sender specifies may have dried up (i.e., trust line limits are exceeded).   

 A rippling transaction can redistribute credit from a more trusted to a less trusted issuer without the specific consent of the involved address' owner~\cite{moreno2016listening}. Gateway and market addresses should allow rippling; however, ordinary addresses may benefit from disallowing rippling on their trust lines.

Since 2013, Ripple has allowed setting a no\_ripple flag to disable rippling transactions involving the trust line. Since March 2015, the flag has been set to true by default; users must opt-in to allow rippling. Furthermore, the defaultRipple flag enables rippling among all the wallet's trust lines.

Researchers in 2016 had found that most paths are short~\cite{moreno2016listening}: 15.1\% of transactions are direct payments. 52.7\% of transactions use a single intermediate address and represent the interactions of gateways with their users following the hot-cold wallet
mechanism. 32\% of transactions use two or more intermediate addresses. In the network, the average path length was 6.83, whereas the diameter was 13.
 
\subsubsection{Financial Concepts}

Ripple adopts and uses the following concepts from traditional finance: offers, checks, and escrows.

\noindent\textbf{Offer.} Offers are orders to exchange two currencies (XRP or issued currencies in any combination). Furthermore, we can trade the same currencies issued by different issuers in offers. An offer is created in an OfferCreate transaction by the sender, and Ripple tries to match the offer by using existing offers fully. If a matching offer is available, the currency balances of both offer senders are updated accordingly. If the offer is not fully consumed (for the requested amount), the offer is written as an offer object to the ledger for the remaining amount. Future transactions can consume the remaining part of the offer. Offers are also used on path-based settlements to convert one currency to another. This auto-bridging of currencies happens automatically on any OfferCreate transaction.

\begin{example} 
An offer $o_2$ that proposes exchanging 10 USD for 7 EUR ($o_2$ sender has 10 USD) will consume an existing offer $o_1$ that proposes exchanging 7 EUR for 10 USD ($o_1$ sender has 7 EUR). Balances of both senders will be updated; $o_2$ sender will have 7 EUR whereas $o_1$ sender will have 10 USD. 

If the exchange rate favors the sender of $o_2$, it will sell less currency. For example, if $o_1$ offers 7 EUR for 9 USD, $o_2$ will take the order and save 1 USD; the sender of $o_2$ buys 7 EUR with 9 USD and keeps its 1 USD. 
\end{example}

Offers are typically used to create cross-currency payments which convert one currency to another by consuming Offers. A cross-currency payment involves two or more currencies; at least one of the currencies must be non-XRP. Cross-currency payments are fully atomic; either the payment fully executes or it fails.

Note that an address can buy user-issued currency through an offer. In that case, the user is deemed to hold an amount of the issued currency, which creates a trust line to the currency issuer. This new trust line is written to the ledger in an object, which increases the reserve XRP requirement of the offer's owner. For this reason, if the owner does not have enough XRP as a reserve, the offer will be considered unfunded and fail.

Offers are explicit instructions to acquire certain issued currencies, so they can go beyond limits set in a trust line.

\noindent\textbf{Check.} A check is a deferred payment that its intended recipient can cash out. Checks record sender, receiver, amount information, and expiration date. The receiver can partially cash a check, and only at the cashing time must the sender have the required amount of the currency in its account. Otherwise, the check fails. Differing from bank checks, both the sender and receiver can cancel a Ripple check before it is cashed out. Checks can send either XRP or issued currencies. 

Checks are helpful for financial institutions to avoid receiving unwanted payments and comply with financial regulations. If a user wants to deposit an amount, it can write a check for a Ripple address of the financial institution and expect the institution to cash it.  The receiving address can fully or partially cash or refuse the check.

Ripple allows rejecting all incoming payments by default, which we can achieve by enabling Deposit Authorization on an address by setting the asDepositAuth flag in an AccountSet transaction. Our address can receive payments from i) pre-authorized addresses or 2) Escrow, Payment Channels, or Checks. 

\noindent\textbf{Escrow.} Escrows are conditional payments that set aside XRP and deliver it when certain conditions are met. The conditions can be time-based unlocks and crypto-conditions. Escrows can expire if not finished in time.

Escrows lock up XRP to be used by a specific receiving address. The escrowed XRP cannot be spent elsewhere nor used or destroyed until the escrow has been successfully finished or canceled. After the expiration time, the unused XRP will return to the sender.

\noindent\textbf{Payment Channel.} Similar to Bitcoin, Ripple uses Payment Channels to set aside XRP that can be used in high-volume microtransactions without recording all channel transactions in ledgers.  

Payment Channels and Escrow can use XRP only, and an address can send XRP to itself through Escrow. However, an address cannot use Payment Channels to send XRP to itself. Checks are more constringent; an address cannot use checks to send XRP nor issued currency to itself.

\noindent\textbf{Partial Payment.} The Amount field of a Payment transaction specifies the exact amount to deliver after charging exchange rates and transfer fees. However, the trust graph may not allow the exact amount to be delivered due to trust line limits, exchange rates, and transfer fees. Rather than failing the transaction altogether, the sender can set the Partial Payment flag (tfPartialPayment) to automatically reduce the starting amount to a deliverable amount. 

Partial payments are notorious because they have been used to exploit naive integrations with the XRP Ledger to steal money from exchanges and gateways.  Assume that an address sends 100 USD to a gateway with the partial payment flag set. The delivered amount (e.g., 30 USD) may be much less than the intended amount. The gateway should not use the amount field to redeem the money. Otherwise, it will pay the address 100 USD and lose 70 USD. Partial Payments have the following limitations: a partial payment cannot provide XRP to satisfy the reserve requirement of an address. Also, direct XRP-to-XRP payments cannot be partial payments.  With all the financial concepts and path-based settlements, 79\% of transactions involve user-issued currencies, whereas 21\% transfer the native cryptocurrency XRP.

\section{IOTA: Tangle Networks}
%In this section, we will give graph models for the IOTA blockchain. We will only cover key aspects that must be known to teach the models. An in-depth analysis of IOTA is given in Chapter~\ref{chapter:iota}.

The IOTA blockchain uses a Proof-of-Work algorithm for mining a transaction, and the concept of a block does not exist. The approach is markedly different from other blockchains, where miners mine blocks of transactions rather than individual transactions.  

We will cover IOTA networks in two sections: the IOTA Tangle graph and the IOTA transaction graph. As a starting point for both graph types, we will discuss the \textit{seed}, \textit{subseed}, private key, difficulty, and hashing concepts that IOTA uses in address and transaction. 

IOTA uses two types of addresses; one-time addresses and Merkle root addresses. One-time addresses send and receive payments. In a sense, one-time addresses are similar to ordinary Bitcoin addresses. Merkle root addresses are used in special applications such as Masked Message Authentication, allowing data channels with publishers and subscribers. 

\begin{remark}
This section will teach one-time addresses because IOTA is phasing out some of the applications that use Merkle root addresses. For example, Masked Message Authentication is being replaced with IOTA Streams. We do not yet know how Merkle root addresses will be used in the new technologies.
\end{remark}

IOTA uses a concept of \textit{seed} which is a string of 81 characters. Each seed character is generated from a tryte, which in turn is three characters (trits) that can take -1, 0, or 1 values. The use of trits, rather than bits that can take 0 or 1, is quite unusual in the blockchain. IOTA defends the practice by arguing that \textquote{ternary computing is considered to be more efficient as it can represent data in three states rather than just two}. For example, $\{0,0,0\}$, $\{0,-1,1\}$ and $\{1,0,1\}$ are three example trytes. As each trit can take three values, a tryte of three trits can take $3^3$ possible values. IOTA uses the number 9 and the uppercase letters A-Z (27 characters in total) to map a tryte to an easy-to-read character. Table~\ref{tab:iota} shows three example trytes and their character mappings. 

\begin{table}
    \centering
    \begin{tabular}{c c c}
     \toprule
        Decimal number&Tryte = 3 Trits&Tryte-encoded character\\
       \midrule
0&	0, 0, 0&9	\\
1&	1, 0, 0	&A\\
2&	-1, 1, 0 &B\\
\bottomrule
    \end{tabular}
    \caption{IOTA uses trytes and encodes each tryte in a character from a 27 character alphabet for ease of use.}
    \label{tab:iota}
\end{table}

A second concept is called an \textit{index}, which can take values between 0 and 9,007,199,254,740,991. We use the index to create multiple addresses from the same seed. The hash value of seed+index is called the \textit{subseed}. IOTA uses the \textit{KERL} hash function, which is the Ternary version of the Keccak-384 hashing algorithm. The hash produces a 243 trit long (81 trytes) subseed. 

\begin{info}
In cryptography, a sponge function or sponge construction is a class of algorithms with a finite internal state that takes an input bitstream of any length and produces an output bitstream of any desired length. Wikipedia.
\end{info}

IOTA uses three security levels to determine the length of the private key: 1, 2, and 3. The private key is computed by absorbing and squeezing the subseed in a \textit{sponge function} 27 times for each security level. The default security level is 2, which produces a private key of 4374 trytes. Longer private keys bring better security; however, they increase the signature length of a transaction (see Figure~\ref{fig:iotaaddresscreation}). 

\begin{figure}
  \begin{center}
    \includegraphics[width=0.45\textwidth]{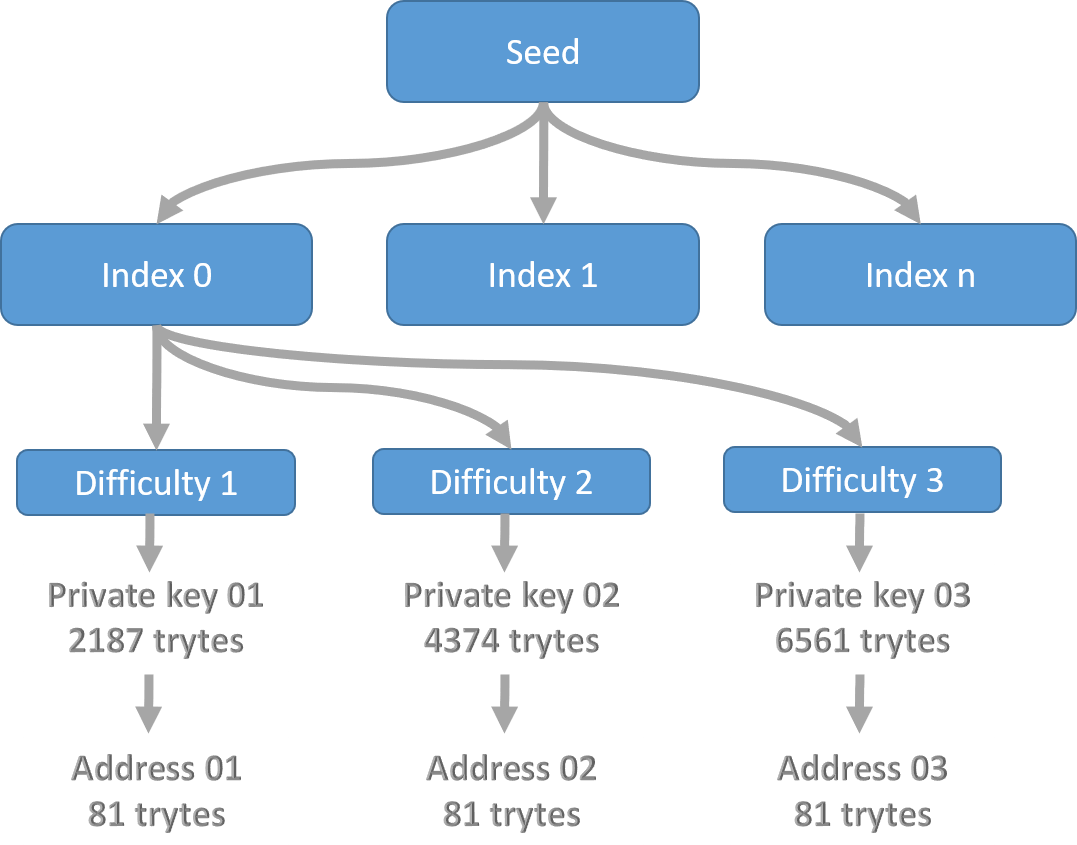}
  \end{center}
  \caption{IOTA address creation by using index and security parameters. An address is created by hashing the computed private key. Note that we compute a different address for each difficulty. A seed can create $9^{57}$ addresses.}
   \label{fig:iotaaddresscreation}
\end{figure}

Signature length causes efficiency problems in transactions. IOTA uses empty transactions that carry neither currency nor a message to store the signature of an address that has security level 2 or 3. Assume that for a security level 2 address $a_i$, we create a (regular) transaction $t_0$ and prepare its signature, which will be 4374 trytes.  However, \textit{signatureMessageFragment} field of a single transaction such as $t_0$ can contain only 2187 trytes. Consequently, we need to create another empty transaction $t_1$ to store the second fragment of the signature. A data structure called \textit{bundle} links transactions $t_0$ and $t_1$  to each other, and the transactions are submitted to the Tangle (i.e., the directed acyclic graph) together. The bundle hash is created from individual transactions of the bundle. As such, any change in any transaction invalidates the bundle hash. For security level 3, we would need to create two empty transactions. Considering that each transaction requires its POW computations and takes up disk storage, the bundling scheme reduces the efficiency of IOTA.  We will cover bundle usages in both the tangle graph and transaction graph. 

\iffalse
\begin{figure}
    \centering
    \includegraphics[width=0.6\linewidth]{figs/iotaaddresscreation.png}
    \caption{IOTA address creation by using index and security parameters. An address is created by hashing the computed private key. Note that we compute a different address for each difficulty. A seed can create $9^{57}$ addresses.}
    \label{fig:iotaaddresscreation}
\end{figure}
\fi

The private key is split into 81-tryte segments where each segment is hashed 26 times. Afterward, segment hashes are hashed together to create an 81 tryte long IOTA address. Sometimes, a checksum of 9 trytes is appended to the address resulting in 90 trytes. An example IOTA address is

\shade{\tiny{OGMMQJUDMNNYSOAXMJWAMNAJPHWMGVAY9UWBXRGTXXVEDIEWSNYRNDQY99NDJQB9QQBPCRRNFAIUPGPLZ}}.

\begin{info}
One-time signatures are public-key signature schemes that have the property that we can use the signatures to sign one single message only. On the other hand, such schemes are usually highly efficient and easy to implement in very constrained devices since they only require the implementation of hash functions and not any advanced arithmetic operations~\cite{dods2005hash}.
\end{info}

 IOTA uses the Winternitz one-time signature scheme to generate digital signatures~\cite{dods2005hash}. Winternitz has the advantage of being robust, easy to compute, and resistant to attacks from quantum computers. On the downside, Winternitz reveals a part of the private key whenever the key signs a transaction. For example, if the address signs two transactions, the signing process reveals more than 50\% of its private key. As a result, an address should sign a single transaction only to minimize risk. The address can receive coins from multiple transactions, but it should spend all of them at one attempt. If the address leaves a balance or receives future payments after it has spent its coins, malicious users can attempt to predict the address private key by using brute-force attacks. As a result, address reuse in IOTA is possible but not recommended. 
\subsection{IOTA tangle graph}
\begin{figure}
  \begin{center}
    \includegraphics[width=0.7\textwidth]{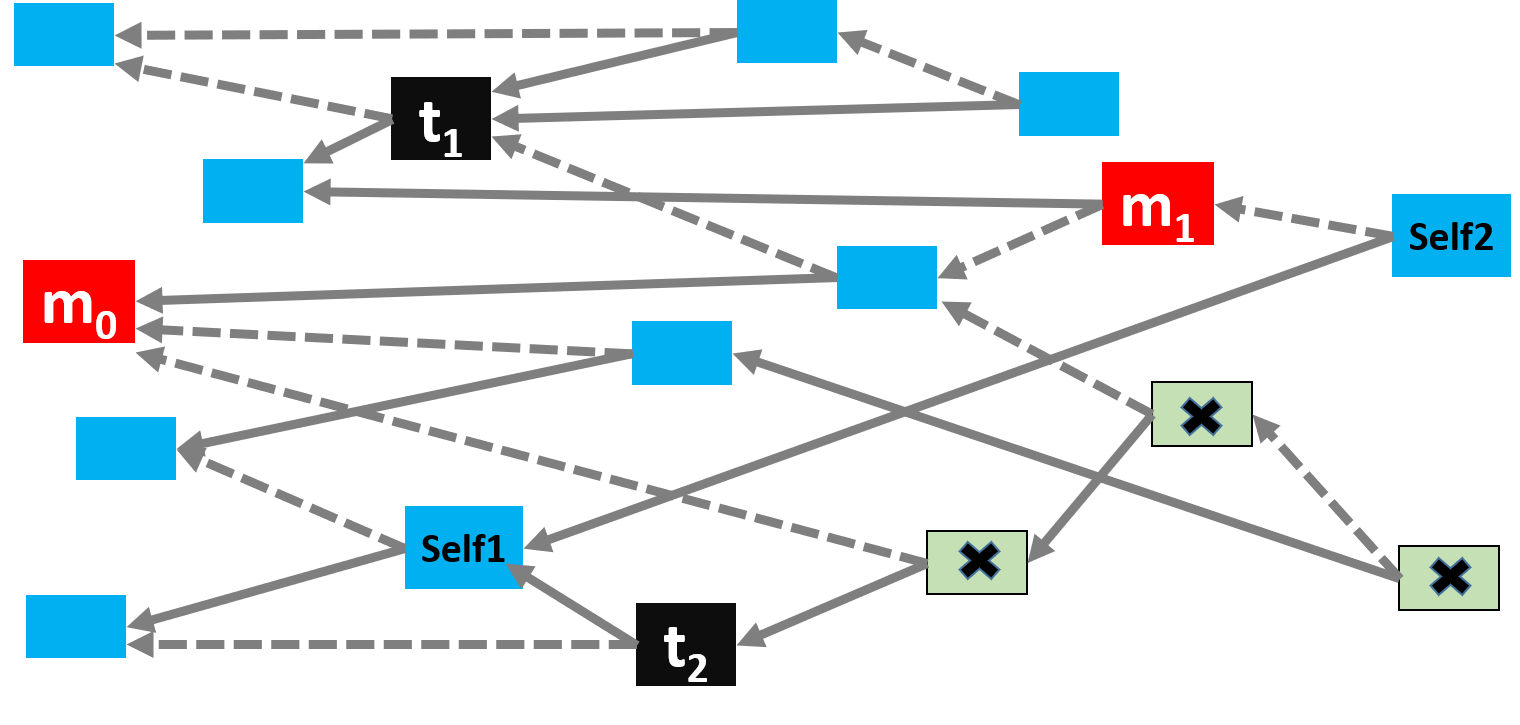}
  \end{center}
  \caption{IOTA tangle graph with milestone transactions $m_0$ and $m_1$. Assume that transactions $t_1$ and $t_2$ double spend the same IOTA coins. As $m_1$ indirectly confirms $t_1$, all transactions (marked with a cross sign) that confirm $t_2$ are building on an invalid subgraph. As a result, these transactions will not be used as tips in future transactions. If a transaction such as Self1 is not confirmed in time, we can create a new zero value transaction Self2 that confirms Self1. New transactions will use Self2 as a tip so that Self1 will be indirectly confirmed by them. Consequently, we hope that Self1 will be indirectly confirmed by a future milestone transaction.}
   \label{fig:tangle}
\end{figure}

An IOTA transaction validates and lists two past transactions, which are called \textit{tips}. Afterward, the transaction is mined with a POW that is considerably easier to solve than the Bitcoin POW. This way, IOTA distributes the mining task to individual users who do not need powerful POW mining devices, which makes IOTA an ideal blockchain for low-powered Internet-of-Thing devices. The resulting transaction dependency network is called the \textit{Tangle}, as shown in  Figure~\ref{fig:tangle}. 

\begin{figure}  
\begin{center}
    \includegraphics[width=0.48\textwidth]{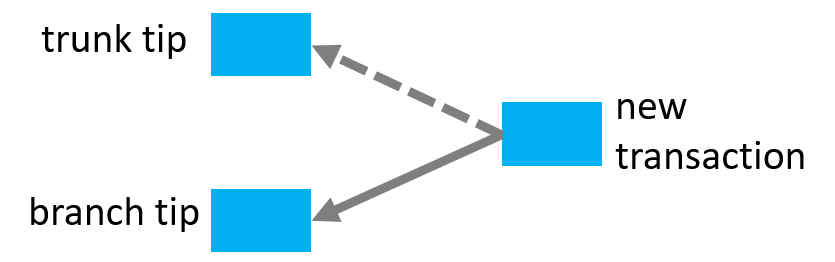}
  \end{center}
  \caption{An IOTA transaction with its two linked past transactions. Edges to tips have dashed and solid fills (as used in official IOTA documentation) depending on their type. Tip types are used to mark information in certain transactions such as in bundles.}
   \label{fig:iotaedge}
\end{figure}

Each new transaction must select two existing transactions and list them as trunk and branch. This process is called tip selection, which is at the heart of IOTA security. Although theoretically, a new transaction could be spending the coins received from one of its tip transactions, the new transaction does not have to have any relation with tips. We can select any previous transaction as a tip.

Multiple new transactions can select an existing transaction as a tip. A transaction $t$ validates its trunk and branch tip directly. Future transactions that use $t$ as a tip confirm the trunk and branch tips of $t$ indirectly. 

Unfortunately, confirmations by individual transactions do not prevent double-spending attacks on IOTA. In Figure~\ref{fig:tangle}, assume that $t_1$ and $t_2$ spend the same IOTA coins in an address $a_1$. Each double-spending transaction can reach a different set of users, who select either $t_1$ or $t_2$ as one of their tips. Note that users would not select both $t_1$ and $t_2$ because they spend the same coins. In traditional blockchains, the block miner would choose to include either $t_1$ or $t_2$, and the network would be safe from the double-spending problem. 

Through an entity known as the \textit{Coordinator}, the IOTA foundation creates and uses \textit{milestone transactions}, which protect the Tangle against double-spending attacks. Milestones arrive every minute and act as anchor points where the confirmed subtangle is considered to be valid and in consensus. The coordinator address is hard-coded in the IOTA software so that all tangle participants can locate and trust milestone transactions. In this aspect, IOTA is a mix of private and public blockchains because although the Tangle is open to the public, the consensus on the Tangle state is guaranteed by the IOTA foundation. In the future, IOTA is aiming to transition to a coordinator-free POW scheme known as \textit{Coordicide} (i.e., killing the coordinator).

In Figure~\ref{fig:tangle}, we show two milestones $m_0$ and $m_1$ which indirectly confirm $t_1$ and invalidate $t_2$. The reason is due to the fact that tangle users will see $t_1$ being confirmed by $m_1$, and the users can mark $t_2$ as double-spending the coins used in $t_1$.

\begin{table}[ht]
    \centering
    \begin{tabular}{c c c c c c c c c c}
    \toprule 
      tx hash&epoch value & value&bundle&tag &address& branch tip & trunk tip\\ \midrule
         
       FPX..999&1603985801 & 0&XTR..QZX&COR..VU& COR..CJZ & EJR..999 & RBL..999\\
       EJR..999&1603985790&0&DKR..NF9&COR..UM&COR..CJZ&MNF..999&NWX..999\\
       RBL..999&1603985788 & 0&PZY..SP9&COR..UL& COR..CJZ&DPW..999&YBD..999\\
         \bottomrule
    \end{tabular}
    \caption{Graph edge list for the Tangle for three zero-value transactions with the same address (we shorten addresses and tx hashes for clarity). A value 0 implies zero-value (message) transactions where the address is the transaction creator. In value transactions, the address may belong to the transaction sender or receiver. The first transaction {\tiny FPXNJVTEVMSAQCUMGRTLGLLNZQOLOUHWXNNYZAXBBEIDXOTTXT9PIRKKBGNTKHPMATIQUHSMNELV99999} has the next two transactions as its tips. Note that these transactions do not belong to the same bundle. The transaction tag (attachmentTag field) is used to label transactions. The epoch value is an integer that counts the number of milliseconds since Jan 1, 1970.}
    \label{tab:tangledata}
\end{table}

\subsection{IOTA transaction graph}

The tangle graph is extended by appending new transactions that confirm tip transactions. We classify IOTA transactions into value and zero-value transactions.

\begin{figure}
    \centering
    \includegraphics[width=0.7\linewidth]{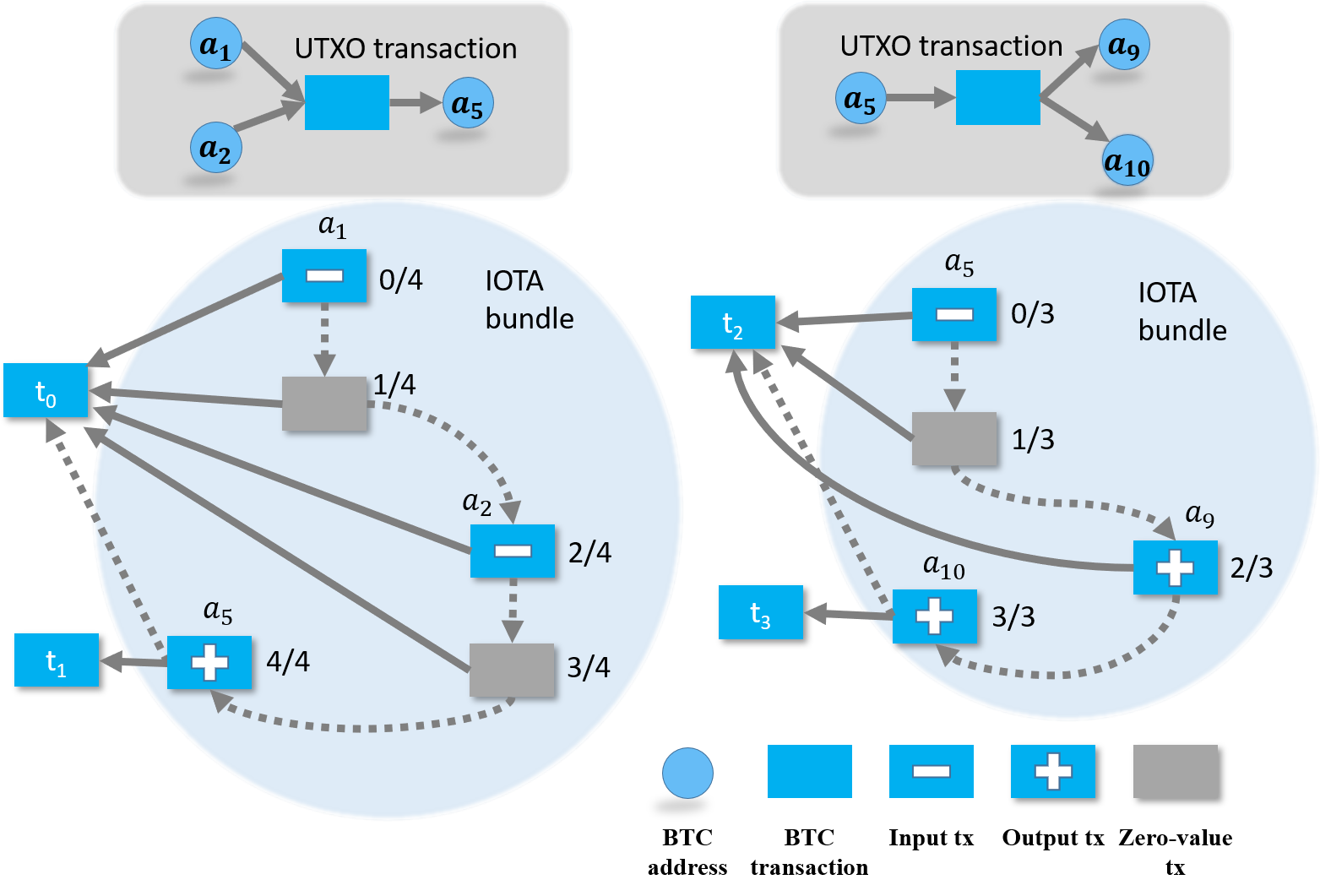}
    \caption{Bundle structure. Two Bitcoin UTXO transactions (top) and their corresponding IOTA bundles (bottom). Dashed edges denote trunk tips, whereas solid edges show branch edges. Tips of bundles ($t_0,\ldots,t_3$) are past transactions that are unrelated to these IOTA bundles. We show trunk tips with dashed edges. The figure assumes that security level=2 for addresses, which requires creating one zero-value transaction (in gray color) to store the second fragment of each input address signature. For the 3rd level, there would be two zero-value transactions for each input address. We give bundle indices (e.g., $0/4$) next to shapes. The left bundle has five transactions, whereas the right bundle has four.}
    \label{fig:iotautxo}
\end{figure}

\noindent\textbf{Zero-value transactions:} IOTA creates a zero-value transaction (see Table~\ref{fig:iotautxo}) for two purposes. First, the transaction may store signature fragments for addresses that use security levels 2 or 3. Second, the transaction may store any data such as sensor data, voting results, or encrypted messages. In both types, data is stored in signatureMessageFragment or attachmentTag fields. All other fields of the transaction are trivial.
 
\noindent\textbf{Value transactions:} IOTA adopts the UTXO transaction model with a few modifications. UTXO model assumes that a transaction has one or more inputs and one or more outputs. For each input, IOTA creates an \textit{input transaction} which lists an address and the withdrawn token value. For each output, IOTA creates an output transaction with an address and a value. The value signs can distinguish input and output transactions: negative for input and positive for output transactions. For the input transaction of an address with security level 1, the address signature is stored in the signatureMessageFragment field of the input transaction. If the address security level is 2 or 3, additional zero-value transactions are created to store second and third key fragments. Output transactions do not require any address signatures.

\begin{table}
    \centering
    \begin{tabular}{c c c c r }
    \toprule 
      transaction&address&time &type&value \\ \midrule
         AXI..99&OOC..H9X &1603951001 &input &-142 998 000\\
         MVP..99&EM9..SYW &1603951006 &output & 1 000\\
         EKG..99&NBN..GJA &1603950941&output& 142 997 000\\
         \bottomrule
    \end{tabular}
    \caption{One input and two outputs transactions from bundle {\tiny SGEAO9WIHW HEEWZRQIVCNIS HBFBE ZCTEYTKTSRUBLXKLD ETKG9ISSUGBOQGFFTDLD GGFBNEIYPYUBEK9C}. IOTA tokens are transferred from the first row address to the next two row addresses. As the input transaction uses security level 2, the bundle contains an additional zero-value transaction (omitted here) to store the second fragment of the signature of OOC..H9X.}
    \label{tab:iotatxdata}
\end{table}

In Table~\ref{tab:iotatxdata} we show a UTXO transaction and its input and output transactions on IOTA. We chose to show the \textquote{type} field to help the reader, but we can as well omit the field because the value can distinguish transaction type. The table omits zero-value transactions that are required to store signatures. In Figure~\ref{fig:iotautxo}, we show two UTXO transactions with all required IOTA bundle transactions. In the figure, note that zero-value transactions are indexed just after their input transactions; $a_1$ signature is stored in index $1/4$. However, the order of the input-output transactions is not important in a bundle; output transaction of address $a_5$ could also be given at index $0/4$. 

As Figure~\ref{fig:iotautxo} clearly shows, an IOTA bundle uses far more pointers and stores many more data pieces than a UTXO transaction. This inefficiency, coupled with low POW difficulty, results in a fast-flowing and large Tangle, which is too costly to store. 

IOTA uses \textit{snapshot}s to limit tangle size periodically. A snapshot resets all transaction history but saves a list of address balances. Only IOTA \textit{permanode}s store all IOTA history. Snapshots occur every few months, and IOTA users are expected to observe the snapshot and manage their IOTA usage accordingly.

\subsection{IOTA Streams}

IOTA Streams is a framework for building cryptographic messaging protocols on IOTA in the Rust programming language. An older version of Streams is known as the Masked Authenticated Messaging, which allowed publishers to attach periodic data (e.g., home temperature readings every hour) as zero value transactions to the Tangle. Streams has a native protocol for attaching messages to the Tangle, but users can extend Streams to send messages from other mediums, such as HTTP URLs.

IOTA Streams allows attaching data in public, private, and subscriber-only modes. In the private case, we can only read Streams messages if the encryption key is known. In the subscriber case, we use a subscription key in addition to the private key.  Streams messages contain a reference for the next (future) message in the Tangle. If the message is encrypted or subscriber-only, we cannot locate the future message in the Tangle. Even if we locate the message, we cannot decrypt its content without the private and subscriber keys. Streams is still in development while its prior-form, the Masked Authentication Messaging, is widely used in the Tangle.

We can model the Streams graph as a hypergraph where nodes are transactions, and directed edges between nodes indicate the shared stream. Note that Streams messages may use previous Streams messages as tips, but this is unnecessary for a well-formed stream.  Streams transactions may be value or zero-value transactions. In the zero-value case, Streams transactions do not have to be confirmed by any other IOTA transaction.

\section{Tools for Blockchain Data Analytics}
\label{sec:tools}

We can access blockchain data easily by downloading wallet software and connecting to the P2P network.  Recently, Google has implemented ETL libraries (\url{https://github.com/blockchain-etl}) to help parse Bitcoin, Monero, ZCash, Ethereum, and a few other blockchains. However, blockchain sizes have become prohibitively large to run data analytics in a single machine. For example, the BlockSci tool~\cite{kalodner2017blocksci} requires around 60GB of memory to load the Bitcoin transaction graph. 

\green{Many blockchain analytics companies offer REST APIs for blockchain data. Examples are etherscan.io, blockcypher.com, and infura.io. However, APIs are either costly or allow limited access to the API. We recommend running your wallet and using the Blockchain-etl libraries of Google to extract blockchain data.}

\green{Most tools and algorithms for blockchain data are related to e-crime or financial (e.g., price, investor) analytics. For example, Barjasic et al~\cite{barjavsic2020time} \textquote{analyze the time-series of minute price returns on the Bitcoin market through the statistical models of the generalized autoregressive conditional heteroskedasticity  family}.  From ransomware payment detection~\cite{akcora2020bitcoinheist} to sextortion discovery~\cite{Egretpaper}, transaction network analysis has proven useful to study blockchain address importance and to cluster blockchain addresses. Network flow algorithms~\cite{WU2019200895}, random walks~\cite{liao2020abnormal} and Petri nets~\cite{wu2020bitcoin} are the main unsupervised methods in this line of research. For example, starting from Bitcoin addresses of potential interest, Egret~\cite{Egretpaper} analyzes neighborhood subgraphs in terms of path length and confluence to detect suspicious Bitcoin flow and other wallet addresses controlled by malicious actors. A rather curious case of clustering research involves identifying heuristics that can link multiple addresses used by a real-life entity. Although heuristics are error-prone, they are widely used for blockchain data analytics (e.g., see~\cite{meiklejohn2013fistful,androulaki2013evaluating,victor2020address})). Another emerging promising approach for ransomware payment detection is topological data analysis (TDA)~\cite{Carlsson, Ghrist}. TDA systematically infers qualitative and quantitative geometric and topological structures of blockchain transaction graphs at multiple resolutions~\cite{Li2019dissecting, abay2019chainnet}. As a result, TDA allows us to capture subtler patterns in the transaction graphs, including changes in chainlet dynamics, which are often associated with illicit or malicious activity and which are inaccessible with more conventional methods based on various forms of information aggregation~\cite{akcora2020bitcoinheist,dorcas2021}.}

\section{Limitations of Data Science on Blockchains}
\label{sec:limitations}
Having covered blockchain data structures, we now turn our attention to the difficulties faced by data scientists when mining blockchain data.

First of all, many exchanges create off-channel transactions and do not store them on the blockchain, and this means that many transactions are executed without being recorded on the blockchain. All of this off-chain transaction information is not directly accessible to data scientists.

 A further concern is to integrate, model, and query data arriving from hundreds of blockchains. The Bitcoin blockchain itself produces approximately 300K transactions per day, arriving from 400K-800K user addresses (see blockchain.com/en/charts), in a temporal resolution scale of 10 minutes. Some low latency blockchains, e.g., IOTA, deliver data in very short intervals. When integrating and querying blockchains, even sifting through the collected data becomes a burden.        
 In almost all cases, Blockchain data exhibit enough quirks and traps that make it challenging to utilize off-the-shelf statistical and Machine Learning techniques.  For example, conventional clustering implies grouping data points that are similar in features.  However, clustering Blockchain data may have a different meaning. In particular, in Blockchain data analytics, clustering ignores similarity and aims to discover which addresses the same user manages. We must develop a completely new suite of algorithms and heuristics for this new definition of clustering. Furthermore, deliberate actions from Blockchain users may render many traditional statistical and data mining methods inefficient. For instance, an aversion to address-reuse in Bitcoin creates transaction networks where address nodes appear two times only (i.e., on receiving and sending bitcoins).
 
On UTXO blockchains, by community practice, addresses are one-time-use only, and address pairs do not have multiple transactions across days. The resulting daily network has more than 400K addresses as nodes, but the median node in-degree is one.  
Even on account base blockchains where addresses are reused, and the coin network (i.e., where the main currency is traded) may be denser, we find that daily token networks (i.e., where tokens are traded) lack community structure. The networks are still very sparse, the clustering coefficient is low, and few communities are formed. Our analysis shows that we remove 99\% of ERC20 token addresses at the first iteration of k-core decomposition.  
Furthermore, events in a global ecosystem impact cryptocurrencies, tokens and assets.  Specifically, Bitcoin significantly impacts the Blockchain ecology; Bitcoin events substantially affect all coin and token networks, and the resulting networks are imbalanced across days. For example, on the Ethereum blockchain, a token such as Storj can have daily transaction numbers ranging from 89 to 34953.

\section{Conclusion}

Despite having their roots in digital money, blockchains rapidly proliferate into virtually every aspect of our everyday life, from enhancing food safety and preventing medical errors to artwork ownership authentication. However,  leveraging the full potential of blockchain technology is impossible without innovative AI tools and algorithms for efficient and systematic analysis of the data stored on the blockchain. Developing such tools, in turn, requires a set of models reflecting the intrinsic blockchain properties for a given application. 

In this tutorial, we have offered a comprehensive overview of the fundamental principles behind the existing blockchain network models (Table~\ref{tab:networktypes}), along with their utility and limitations across a wide range of blockchain applications and blockchain data analytic tasks.

\begin{table}
    \centering
    \begin{tabular}{l c c c c}
    
    & Weighted &  Directed & Multigraph & Hypergraph\\
    \midrule
       Bitcoin address graph& \checkmark& \checkmark&\checkmark&\\
      \rowcolor{Gray} Bitcoin transaction graph&\checkmark &\checkmark &&\\
       Ethereum transaction graph& \checkmark & \checkmark &\checkmark&\\
     \rowcolor{Gray} Ethereum token graph& \checkmark & \checkmark &\checkmark&\\
       Ethereum trace graph&  &  \checkmark &&\checkmark\\
     \rowcolor{Gray} Ripple trust graph& \checkmark&\checkmark &\checkmark&\\
      Ripple payment graph& \checkmark&\checkmark &&\checkmark\\
     \rowcolor{Gray} IOTA tangle graph& & \checkmark&&\\
      IOTA transaction graph&\checkmark &\checkmark &&\\
     \rowcolor{Gray} IOTA stream graph& &\checkmark &&\checkmark\\
         \bottomrule
    \end{tabular}
    \caption{Graph characteristics of blockchain networks. }
    \label{tab:networktypes}
\end{table}

\bibliography{wiley}

\begin{thebibliography}{34}
\providecommand{\natexlab}[1]{#1}
\providecommand{\url}[1]{\texttt{#1}}
\providecommand{\urlprefix}{}

\bibitem[{Nakamoto(2008)Nakamoto, Satoshi}]{nakamoto2008bitcoin}
Nakamoto S, Satoshi, editor, Bitcoin: A peer-to-peer electronic cash system.
\newblock bitcoin.org; 2008.

\bibitem[{Dwork and Naor(1992)Dwork, Cynthia and Naor, Moni}]{dwork1992pricing}
Dwork C, Naor M.
\newblock Pricing via processing or combatting junk mail.
\newblock In: Annual International Cryptology Conference Springer; 1992. p.
  139--147.

\bibitem[{Cachin et~al.(2016)Cachin, Christian and
  others}]{cachin2016architecture}
Cachin C, et~al.
\newblock Architecture of the hyperledger blockchain fabric.
\newblock In: Workshop on distributed cryptocurrencies and consensus ledgers,
  vol. 310; 2016. p. 1--12.

\bibitem[{Szabo(1997)Szabo, Nick}]{szabo1997idea}
Szabo N.
\newblock The idea of smart contracts.
\newblock Nick Szabo’s Papers and Concise Tutorials 1997;6.

\bibitem[{Poon and Dryja(2016)Poon, Joseph and Dryja,
  Thaddeus}]{poon2016bitcoin}
Poon J, Dryja T, Poon, editor, The bitcoin lightning network: Scalable
  off-chain instant payments.
\newblock Nakamoto Institute; 2016.
\newblock
  \urlprefix\url{https://nakamotoinstitute.org/static/docs/lightning-network.pdf}.

\bibitem[{Ruffing et~al.(2014)Ruffing, Tim and Moreno-Sanchez, Pedro and Kate,
  Aniket}]{ruffing2014coinshuffle}
Ruffing T, Moreno-Sanchez P, Kate A.
\newblock CoinShuffle: Practical decentralized coin mixing for Bitcoin.
\newblock In: European Symposium on Research in Computer Security Springer;
  2014. p. 345--364.

\bibitem[{Milo et~al.(2002)Milo, Ron and Shen-Orr, Shai and Itzkovitz, Shalev
  and Kashtan, Nadav and Chklovskii, Dmitri and Alon, Uri}]{milo2002network}
Milo R, Shen-Orr S, Itzkovitz S, Kashtan N, Chklovskii D, Alon U.
\newblock Network motifs: simple building blocks of complex networks.
\newblock Science 2002;298(5594):824--827.

\bibitem[{Noether(2015)Noether, Shen}]{noether2015ring}
Noether S.
\newblock Ring SIgnature Confidential Transactions for Monero.
\newblock IACR Cryptol ePrint Arch 2015;2015:1098.

\bibitem[{Kumar et~al.(2017)Kumar, Amrit and Fischer, Cl{\'e}ment and Tople,
  Shruti and Saxena, Prateek}]{kumar2017traceability}
Kumar A, Fischer C, Tople S, Saxena P.
\newblock A traceability analysis of Monero’s blockchain.
\newblock In: European Symposium on Research in Computer Security Springer;
  2017. p. 153--173.

\bibitem[{M{\"o}ser et~al.(2018)M{\"o}ser, Malte and Soska, Kyle and Heilman,
  Ethan and Lee, Kevin and Heffan, Henry and Srivastava, Shashvat and Hogan,
  Kyle and Hennessey, Jason and Miller, Andrew and Narayanan, Arvind and
  others}]{moser2018empirical}
M{\"o}ser M, Soska K, Heilman E, Lee K, Heffan H, Srivastava S, et~al.
\newblock An empirical analysis of traceability in the monero blockchain.
\newblock Proceedings on Privacy Enhancing Technologies 2018;2018(3):143--163.

\bibitem[{Alonso(2018)Alonso, Kurt Magnus}]{alonso2018monero}
Alonso KM.
\newblock Monero-privacy in the blockchain.
\newblock University Press 2018;.

\bibitem[{Biryukov and Tikhomirov(2019)Biryukov, Alex and Tikhomirov,
  Sergei}]{biryukov2019deanonymization}
Biryukov A, Tikhomirov S.
\newblock Deanonymization and linkability of cryptocurrency transactions based
  on network analysis.
\newblock In: 2019 IEEE European Symposium on Security and Privacy (EuroS\&P)
  IEEE; 2019. p. 172--184.

\bibitem[{Akcora et~al.(2018{\natexlab{a}})Akcora, Cuneyt Gurcan and Dey, Asim
  K. and Gel, Yulia R. and Kantarcioglu, Murat}]{akcora2017chainlet}
Akcora CG, Dey AK, Gel YR, Kantarcioglu M.
\newblock Forecasting Bitcoin Price with Graph Chainlets.
\newblock In: The Pacific-Asia Conference on Knowledge Discovery and Data
  Mining (PAKDD), Melbourne, Australia; 2018. p. 1--12.

\bibitem[{Akcora et~al.(2018{\natexlab{b}})Akcora, Cuneyt Gurcan and Dixon,
  Matthew and Gel, Yulia R. and Kantarcioglu, Murat}]{akcora2018EL}
Akcora CG, Dixon M, Gel YR, Kantarcioglu M.
\newblock Bitcoin Risk Modeling with Blockchain Graphs.
\newblock Economics Letters 2018;p. 1--5.

\bibitem[{Dey et~al.(2020)Dey, Asim Kumar and Akcora, Cuneyt Gurcan and Gel,
  Yulia R. and Kantarcioglu, Murat}]{deyCSDA2018}
Dey AK, Akcora CG, Gel YR, Kantarcioglu M.
\newblock On the Role of Local Blockchain Network Features in Cryptocurrency
  Price Formation.
\newblock Canadian Journal of Statistics 2020;p. 1--35.

\bibitem[{Lee et~al.(2020)Lee, Xi Tong and Khan, Arijit and Sen Gupta, Sourav
  and Ong, Yu Hann and Liu, Xuan}]{lee2019measurements}
Lee XT, Khan A, Sen~Gupta S, Ong YH, Liu X.
\newblock Measurements, Analyses, and Insights on the Entire Ethereum
  Blockchain Network.
\newblock In: Proceedings of The Web Conference 2020 WWW'20, New York, NY, USA:
  Association for Computing Machinery; 2020. p. 155--166.
\newblock \urlprefix\url{https://doi.org/10.1145/3366423.3380103}.

\bibitem[{El~Qorchi et~al.(2003)El Qorchi, Mohammed and Maimbo, Samuel Munzele
  and Autmainbo, Samuel Munzele and Wilson, John F and others}]{el2003informal}
El~Qorchi M, Maimbo SM, Autmainbo SM, Wilson JF, et~al.
\newblock Informal Funds Transfer Systems: An analysis of the informal hawala
  system, vol. 222.
\newblock International Monetary Fund; 2003.

\bibitem[{Moreno-Sanchez et~al.(2016)Moreno-Sanchez, Pedro and Zafar, Muhammad
  Bilal and Kate, Aniket}]{moreno2016listening}
Moreno-Sanchez P, Zafar MB, Kate A.
\newblock Listening to whispers of ripple: Linking wallets and deanonymizing
  transactions in the ripple network.
\newblock Proceedings on Privacy Enhancing Technologies 2016;2016(4):436--453.

\bibitem[{Dods et~al.(2005)Dods, Chris and Smart, Nigel P and Stam,
  Martijn}]{dods2005hash}
Dods C, Smart NP, Stam M.
\newblock Hash based digital signature schemes.
\newblock In: IMA International Conference on Cryptography and Coding Springer;
  2005. p. 96--115.

\bibitem[{Kalodner et~al.(2017)Kalodner, Harry and Goldfeder, Steven and
  Chator, Alishah and M{\"o}ser, Malte and Narayanan,
  Arvind}]{kalodner2017blocksci}
Kalodner H, Goldfeder S, Chator A, M{\"o}ser M, Narayanan A.
\newblock BlockSci: Design and applications of a blockchain analysis platform.
\newblock arXiv preprint arXiv:170902489 2017;.

\bibitem[{Barja{\v{s}}i{\'c} and Antulov-Fantulin(2020)Barja{\v{s}}i{\'c},
  Irena and Antulov-Fantulin, Nino}]{barjavsic2020time}
Barja{\v{s}}i{\'c} I, Antulov-Fantulin N.
\newblock Time-varying volatility in Bitcoin market and information flow at
  minute-level frequency.
\newblock arXiv preprint arXiv:200400550 2020;.

\bibitem[{Akcora et~al.(2020)Akcora, Cuneyt G and Li, Yitao and Gel, Yulia R
  and Kantarcioglu, Murat}]{akcora2020bitcoinheist}
Akcora CG, Li Y, Gel YR, Kantarcioglu M.
\newblock Bitcoinheist: Topological data analysis for ransomware prediction on
  the bitcoin blockchain.
\newblock In: Proceedings of the twenty-ninth international joint conference on
  artificial intelligence; 2020. p. 1--7.

\bibitem[{S.~Phetsouvanh(2018)S. Phetsouvanh, F. Oggier, A. Datta}]{Egretpaper}
S~Phetsouvanh AD F~Oggier.
\newblock EGRET: Extortion Graph Exploration Techniques in the Bitcoin Network.
\newblock In: ICDM Workshop on Data Mining in Networks (DaMNet) IEEE; 2018. p.
  1--6.

\bibitem[{Wu et~al.(2019)Yan Wu and Anthony Luo and Dianxiang
  Xu}]{WU2019200895}
Wu Y, Luo A, Xu D.
\newblock Identifying suspicious addresses in Bitcoin thefts.
\newblock Digital Investigation 2019;31:200895.
\newblock
  \urlprefix\url{https://www.sciencedirect.com/science/article/pii/S1742287619302233}.

\bibitem[{Liao et~al.(2020)Liao, Qian and Gu, Yijun and Liao, Junfan and Li,
  Wenzheng}]{liao2020abnormal}
Liao Q, Gu Y, Liao J, Li W.
\newblock Abnormal transaction detection of Bitcoin network based on feature
  fusion.
\newblock In: 2020 IEEE 9th Joint International Information Technology and
  Artificial Intelligence Conference (ITAIC), vol.~9 IEEE; 2020. p. 542--549.

\bibitem[{Wu et~al.(2020)Wu, Yan and Tao, Fang and Liu, Lu and Gu, Jiayan and
  Panneerselvam, John and Zhu, Rongbo and Shahzad, Mohammad
  Nasir}]{wu2020bitcoin}
Wu Y, Tao F, Liu L, Gu J, Panneerselvam J, Zhu R, et~al.
\newblock A bitcoin transaction network analytic method for future blockchain
  forensic investigation.
\newblock IEEE Transactions on Network Science and Engineering 2020;.

\bibitem[{Meiklejohn et~al.(2013)Meiklejohn, S. and Pomarole, M. and Jordan, G.
  and Levchenko, K.and McCoy, D. and Voelker, G. M and Savage,
  S.}]{meiklejohn2013fistful}
Meiklejohn S, Pomarole M, Jordan G, Levchenko K, McCoy D, Voelker GM, et~al.
\newblock A fistful of bitcoins: characterizing payments among men with no
  names.
\newblock In: Proceedings of the 2013 conference on Internet measurement
  conference ACM; 2013. p. 127--140.

\bibitem[{Androulaki et~al.(2013)Androulaki, Elli and Karame, Ghassan O and
  Roeschlin, Marc and Scherer, Tobias and Capkun,
  Srdjan}]{androulaki2013evaluating}
Androulaki E, Karame GO, Roeschlin M, Scherer T, Capkun S.
\newblock Evaluating user privacy in bitcoin.
\newblock In: IFCA Springer; 2013. p. 34--51.

\bibitem[{Victor(2020)Victor, Friedhelm}]{victor2020address}
Victor F.
\newblock Address clustering heuristics for Ethereum.
\newblock In: International Conference on Financial Cryptography and Data
  Security Springer; 2020. p. 617--633.

\bibitem[{Carlsson(2009)Carlsson, Gunnar}]{Carlsson}
Carlsson G.
\newblock Topology and data.
\newblock BAMS 2009;46(2):255--308.

\bibitem[{Ghrist(2008)Ghrist, R.}]{Ghrist}
Ghrist R.
\newblock Barcodes: The persistent topology of data.
\newblock BAMS 2008;45(1):61--75.

\bibitem[{Li et~al.(2019)Li, Yitao and Islambekov, Umar and Akcora, Cuneyt and
  Smirnova, Ekaterina and Gel, Yulia R and Kantarcioglu,
  Murat}]{Li2019dissecting}
Li Y, Islambekov U, Akcora C, Smirnova E, Gel YR, Kantarcioglu M.
\newblock Dissecting Ethereum Blockchain Analytics: What We Learn from Topology
  and Geometry of Ethereum Graph.
\newblock arXiv preprint arXiv:191210105 2019;.

\bibitem[{Abay et~al.(2019)Abay, Nazmiye Ceren and Akcora, Cuneyt Gurcan and
  Gel, Yulia R and Kantarcioglu, Murat and Islambekov, Umar D and Tian, Yahui
  and Thuraisingham, Bhavani}]{abay2019chainnet}
Abay NC, Akcora CG, Gel YR, Kantarcioglu M, Islambekov UD, Tian Y, et~al.
\newblock Chainnet: Learning on blockchain graphs with topological features.
\newblock In: ICDM; 2019. p. 946--951.

\bibitem[{Ofori-Boateng et~al.(2021)Ofori-Boateng, Dorcas and Segovia
  Dominguez, Ignacio Jesus and Akcora, Cuneyt Gurcan and and Kantarcioglu,
  Murat and Gel, Yulia R}]{dorcas2021}
Ofori-Boateng D, Segovia~Dominguez IJ, Akcora CG, , Kantarcioglu M, Gel YR.
\newblock Topological anomaly detection in dynamic multilayer blockchain
  networks.
\newblock In: ECML; 2021. p. 1--9.

\end{thebibliography}
\end{document}